\documentclass[acmsmall,screen,nonacm]{acmart}

\usepackage{pifont}
\usepackage{listings}
\usepackage{placeins}
\usepackage{wrapfig}
\usepackage{enumitem}
\setlist[itemize]{leftmargin=*}
\setlist[enumerate]{leftmargin=*}
\usepackage{tikz}
\usetikzlibrary{arrows.meta}
\usepackage{pgfplots}
\pgfplotsset{compat=1.18}

\makeatletter
\renewcommand{\@parfont}{\normalfont\bfseries}
\makeatother

\makeatletter
\let\algorithmruledstyle\fs@ruled
\renewcommand{\fs@ruled}{%
  \algorithmruledstyle
  \def\@fs@post{\kern2pt\hrule\kern-6pt}}
\makeatother
\floatstyle{ruled}
\newfloat{algorithm}{tbp}{loa}
\floatname{algorithm}{Algorithm}
\floatstyle{plain}

\definecolor{thupurple}{HTML}{660874}  %
\definecolor{figfillA}{HTML}{EEEAF7}   %
\definecolor{figfillB}{HTML}{F6EBF2}   %
\definecolor{figfillC}{HTML}{F8F2E3}   %
\definecolor{figfillN}{HTML}{F4F3F6}   %
\definecolor{figgridA}{HTML}{B8A9DC}   %
\definecolor{figgridB}{HTML}{D3A5C9}   %

\lstdefinestyle{leanexample}{
  basicstyle=\ttfamily\footnotesize,
  keywordstyle=\color{thupurple}\bfseries,
  morekeywords={abbrev,def,fun},
  morekeywords=[2]{Type,Nat,Event,TraceInf},
  keywordstyle=[2]\color{blue!55!black},
  sensitive=true,
  columns=fullflexible,
  keepspaces=true,
  showstringspaces=false,
  frame=single,
  rulecolor=\color{figgridA},
  backgroundcolor=\color{figfillN},
  framerule=0.3pt,
  framesep=5pt,
  xleftmargin=6pt,
  xrightmargin=6pt,
  aboveskip=6pt,
  belowskip=6pt,
  breaklines=true
}

\newcommand{\figfont}{\fontfamily{ptm}\selectfont}

\definecolor{posgreen}{HTML}{2E7D32}
\definecolor{negred}{HTML}{C62828}

\DeclareRobustCommand{\chen}[1]{{\color{black}{}#1}}

\begin{document}

\title{CompCert4Lean: Porting a Verified C Compiler to Lean\texorpdfstring{\\}{ }in Six Weeks with Parallel Coding Agents}

\author{Cong Wang}
\affiliation{\institution{Tsinghua University}\city{Beijing}\country{China}}
\email{wcon006@gmail.com}

\author{Zhiyang Chen}
\affiliation{\institution{University of Toronto}\city{Toronto}\country{Canada}}
\email{zhiychen@cs.toronto.edu}

\author{Fan Long}
\affiliation{\institution{University of Toronto}\city{Toronto}\country{Canada}}
\email{fanl@cs.toronto.edu}

\author{Martin Rinard}
\affiliation{\institution{Massachusetts Institute of Technology}\city{Cambridge}\country{USA}}
\email{rinard@csail.mit.edu}

\begin{abstract}
We present CompCert4Lean, a large-scale port of CompCert~3.17's verified
compiler core from Rocq to Lean, completed in approximately six weeks by
one expert directing parallel LLM agents. Our migration workflow
partitions the development along its dependencies, audits translated
definitions and theorem statements before large-scale proof
reconstruction, and then reconstructs proofs in parallel against stable
interfaces. The resulting compiler covers typed CompCert C through
x86-64 assembly, discharges all 4,908 recorded proof obligations without
project-specific axioms, and produces executable Linux binaries. The
workflow enabled the expert to concentrate on organizing the work rather
than supplying technical solutions. Of 1,071 human interactions, 88.8\%
concerned workflow and coordination, while only three supplied technical
direction.

Optimization campaigns further demonstrate the benefit of the Lean port
for agent-driven verified development. Under matched budgets of 24 active
hours, GPT-6-astra produced a Lean candidate with 11.7\% lower
generated-program runtime across 29 small-tests programs, compared with
2.3\% lower for its Rocq candidate, both relative to original
CompCert~3.17 and while preserving the end-to-end correctness theorem.
These results show that parallel LLM agents can support large-scale
migration with limited human technical direction and subsequently improve
the migrated compiler while retaining its formal guarantees.
\end{abstract}



\maketitle

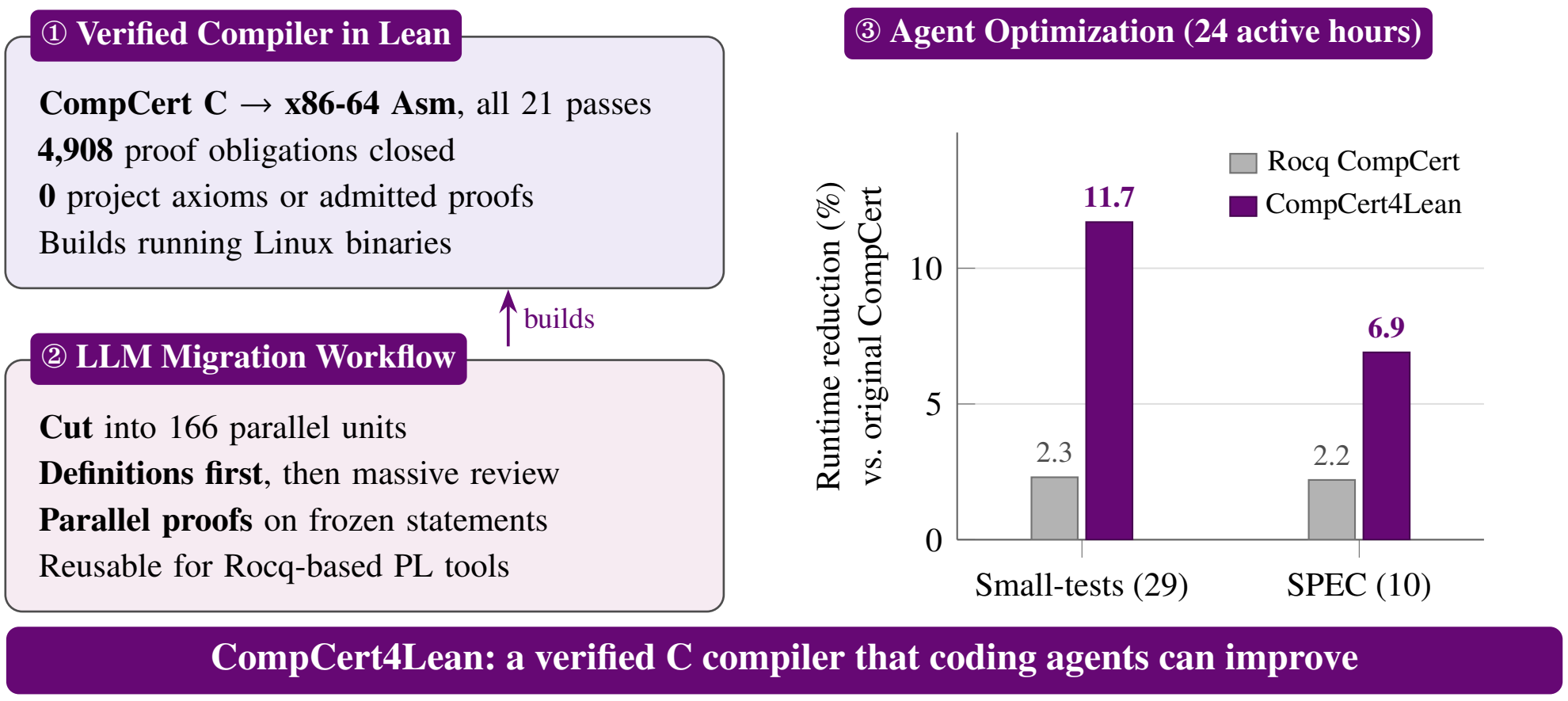
\begin{figure}[t]
  \centering
  \begin{tikzpicture}[
    font=\figfont,
    /pgf/number format/assume math mode=true,
    idea/.style={draw=black!70, line width=0.5pt, rounded corners=4pt,
      fill=#1, minimum width=60mm, minimum height=21mm},
    tab/.style={fill=thupurple, text=white, font=\figfont\bfseries\footnotesize,
      rounded corners=2pt, inner sep=2.5pt},
    bullet/.style={align=left, font=\figfont\footnotesize, text width=55mm,
      anchor=north west, inner sep=0pt},
  ]
    \node[idea=figfillA] (c) at (0,1.15) {};
    \node[tab, anchor=west] at ([xshift=6pt]c.north west)
      {\ding{172}\ Verified Compiler in Lean};
    \node[bullet] at ([xshift=8pt,yshift=-13pt]c.north west) {%
      \textbf{CompCert C $\rightarrow$ x86-64 Asm}, all 21 passes\\[1pt]
      \textbf{4,908} proof obligations closed\\[1pt]
      \textbf{0} project axioms or admitted proofs\\[1pt]
      Builds running Linux binaries};

    \node[idea=figfillB] (m) at (0,-1.55) {};
    \node[tab, anchor=west] at ([xshift=6pt]m.north west)
      {\ding{173}\ LLM Migration Workflow};
    \node[bullet] at ([xshift=8pt,yshift=-13pt]m.north west) {%
      \textbf{Cut} into 166 parallel units\\[1pt]
      \textbf{Definitions first}, then massive review\\[1pt]
      \textbf{Parallel proofs} on frozen statements\\[1pt]
      Reusable for \chen{Rocq}-based PL tools};

    \draw[-{Stealth[length=2mm]}, line width=0.7pt, thupurple]
      ([xshift=12mm,yshift=3pt]m.north) -- node[right, font=\figfont\scriptsize, text=thupurple]
      {builds} ([xshift=12mm]c.south);

    \node[tab, anchor=north west] at (4.0,2.45) {\ding{174}\ Agent Optimization (24 active hours)};
    \begin{axis}[
      at={(4.95cm,-2.0cm)}, anchor=south west,
      scale only axis, width=44mm, height=34mm,
      ybar, bar width=11pt,
      ymin=0, ymax=15,
      enlarge x limits=0.45,
      symbolic x coords={C, SPEC},
      xticklabels={{Small-tests (29)}, {SPEC (10)}},
      xtick=data, ytick={0,5,10},
      ylabel={\shortstack{Runtime reduction (\%)\\vs. original CompCert}},
      ylabel style={font=\figfont\footnotesize, yshift=-2pt},
      ticklabel style={font=\figfont\footnotesize},
      yticklabel={\pgfmathprintnumber{\tick}},
      axis line style={black!60}, tick style={black!60},
      ymajorgrids, grid style={black!12},
      axis x line*=bottom, axis y line*=left,
      nodes near coords={\pgfmathprintnumber[fixed,fixed zerofill,precision=1]{\pgfplotspointmeta}},
      every node near coord/.append style={font=\figfont\scriptsize, text=black!70},
      legend style={draw=none, fill=none, font=\figfont\scriptsize,
        at={(1.0,1.0)}, anchor=north east, legend columns=1},
      legend image code/.code={\draw[#1] (0cm,-0.08cm) rectangle (0.22cm,0.08cm);},
    ]
      \addplot[fill=black!30, draw=black!55] coordinates {(C,2.3) (SPEC,2.2)};
      \addplot[fill=thupurple, draw=thupurple!70!black,
        every node near coord/.append style={text=thupurple, font=\figfont\bfseries\scriptsize}]
        coordinates {(C,11.7) (SPEC,6.9)};
      \legend{Rocq CompCert, CompCert4Lean}
    \end{axis}

    \node[fill=thupurple, text=white, font=\figfont\bfseries\small,
      rounded corners=3pt, minimum width=130mm, inner sep=4pt, anchor=north west]
      at (-3.0,-2.72)
      {CompCert4Lean: a verified C compiler that coding agents can improve};
  \end{tikzpicture}
  \vspace{-6pt}
  \caption{CompCert4Lean: LLM-agent migration and verified optimization.
  \ding{172} Full CompCert C-to-x86-64 compilation; all 4,908 proof obligations closed.
  \ding{173} Definitions-first migration across 166 units, large-scale review,
  then parallel proof reconstruction. \ding{174} With 24 active hours each,
  GPT-6-astra reduces runtime by 11.7\% on Lean versus 2.3\% on \chen{Rocq}
  across 29 small-tests programs, and 6.9\% versus 2.2\% across ten SPEC inputs,
  relative to original CompCert~3.17. Both retain the end-to-end correctness theorem
  (Section~\ref{sec:evaluation}).}

  \Description{Left: two boxes summarize the verified Lean compiler and the
  LLM migration workflow. Right: a bar chart after 24 active hours of optimization, showing runtime reduction relative
  to original CompCert 3.17. CompCert4Lean reaches 11.7\% across 29 small-tests
  programs and 6.9\% across ten SPEC inputs, versus 2.3\% and 2.2\% for
  the selected Rocq CompCert candidate.}
  \label{fig:intro:teaser}
  \vspace{-10pt}
\end{figure}
 
\section{Introduction}
\label{sec:introduction}

\chen{Lean has made substantial progress as a platform for formalized
mathematics and automated theorem proving~\cite{demoura2021lean4}. Its
Mathlib library~\cite{mathlib2020} provides an extensive foundation of
reusable mathematical definitions and theorems. AI systems are now using
this infrastructure to tackle long-standing open research problems. In
October 2026, OpenAI released more than 700 mathematical manuscripts
reporting solutions and advances across number theory, geometry,
analysis, and theoretical computer science~\cite{openai2026math}. Among
these is a reported proof of the quasi-Riemann hypothesis, a weaker form
of the celebrated Riemann hypothesis, accompanied by a Lean
formalization~\cite{openai2026quasiriemann}.
Many other results in the collection also include formal proofs that
can be checked mechanically~\cite{openai2026math}. These advances create
an opportunity to bring increasingly capable proof agents and extensive
mathematical libraries to the construction, maintenance, and extension
of verified software systems.}

\chen{Despite Lean's progress, a substantial body of mature
program-verification infrastructure remains available only in
Rocq~\cite{rocq2025}. A prominent example is CompCert, an optimizing C
compiler with a machine-checked proof of semantic preservation from C to
assembly~\cite{leroy2009formal,leroy2009backend}.
Its semantics, memory model, and correctness proofs also support Rocq
developments in relaxed-memory concurrency~\cite{sevcik2013compcerttso},
separate and compositional
compilation~\cite{stewart2015compositional,song2020compcertm,koenig2021compcerto},
C program logics~\cite{appel2014vst}, static
analysis~\cite{jourdan2015verasco}, and verified operating-system
kernels~\cite{gu2016certikos}. This infrastructure embodies years of
proof engineering, yet its definitions and theorems remain tied to Rocq
and cannot be imported directly into Lean.}

\chen{Bringing CompCert and similar infrastructure to Lean would make existing
verification work reusable within the Lean ecosystem. Other developments
could build on its language semantics, memory model, and correctness
theorems without reconstructing those foundations. The growing capability
of agents to construct Lean proofs~\cite{openai2026math} also creates an
opportunity to reduce the effort of maintaining and extending verified
software. For example, an agent could add a compiler optimization together
with the proof needed to preserve the end-to-end correctness theorem.
Migration would thus enable new developments to reuse established
verification work while making that infrastructure more accessible to
current proof agents.}

\textbf{CompCert4Lean.}
\chen{To realize these benefits, we present CompCert4Lean, a Lean port of the
CompCert~3.17 compiler core. One author directed the main migration development with parallel LLM
coding agents, completing the port in approximately six weeks. The port
discharges all 4,908 recorded proof obligations without project-specific
axioms and produces executable Linux binaries. The port also pays off
immediately. Even though Rocq CompCert is mature, documented, and
publicly available, a coding agent under the same budget of 24 active hours
produces a Lean candidate with 11.7\% lower generated-program runtime
across 29 small-tests programs, compared with 2.3\% lower for its Rocq
candidate, both relative to original CompCert~3.17.}

\chen{Achieving this migration in six weeks required overcoming substantial
proof-engineering challenges. Even the first verified version of CompCert
consisted of 42,000 lines of Rocq and took about three person-years
over five years~\cite{leroy2009formal}. In CompCert~3.17, the 154 source
modules corresponding to our migration scope contain 149,038 lines
of Rocq. Porting this development requires reconstructing executable
code, language semantics, theorem statements, and proofs while
preserving their relationships. Its proof scripts depend on Rocq's
tactics, libraries such as Flocq~\cite{boldo2011flocq}, and details
of its logic. Differences between Rocq and Lean can require new
representations and proof strategies, and a change to a shared
definition can affect many downstream proofs.}

Addressing these challenges with coding agents introduces a further
difficulty: agent throughput depends on parallelism, whereas a verified
development is a dense web of dependencies among definitions, statements,
and proofs. Simply assigning different modules
to concurrent agents does not make their work independent. A mistake in
a shared definition or theorem statement can become the basis for many
agents' work, spreading into further definitions, interfaces, and proofs.
Repairing that mistake later can require changes across multiple units
and invalidate substantial proof effort. Naive parallelism can therefore
amplify errors as well as accelerate progress. Successful builds and
kernel-checked proofs do not eliminate this risk: Lean can check a proof
of an incorrectly translated proposition without establishing that it
preserves the intended source statement. Section~\ref{ex:constructor-interface}
illustrates this interface risk with an inductive-interface repair.

Our workflow (Figure~\ref{fig:migration-workflow}) addresses these challenges
by combining dependency-based partitioning with review before large-scale
proof reconstruction. We partition the formalization into units with
canonical owners and stable shared interfaces, allowing agents to work
in parallel once their dependencies are available. For CompCert, this
yields 166 units: 154 source components and 12 shared providers. Agents
first translate definitions and exact theorem statements, completing
necessary supporting proofs and recording the remaining duties as
explicit obligations. Independent reviewers then audit the
translation against the \chen{Rocq} source, and review and repair iterate
before the baseline is accepted for proof reconstruction. 
Before large-scale proof reconstruction, the first audit caught 1,091
theorem statements reduced to \texttt{True}, exposing errors that
translation-time builds had missed.
Partitioning organizes concurrent work around consistent interfaces,
while early review reduces the risk that errors in those interfaces and
statements propagate into a large body of downstream proofs. Agents then
reconstruct proofs against the accepted definitions and
statements.

\textbf{Experimental Results.}
Our migration produces a verified compiler core comprising 173 Lean
modules and discharges all 4,908 recorded proof obligations. Its
end-to-end theorem establishes semantic preservation from typed
CompCert C abstract syntax to x86-64 assembly abstract syntax. A
transitive audit of 75,348 project declarations finds no axioms beyond
Lean's three standard foundational axioms. The executable toolchain
passes all 139 applicable CompCert small tests at both O0 and O2,
together with 20 correctness executions across six SPEC CPU2026 programs
and the two optimization levels. Independent source reviews and
differential tests\chen{---including identical RTL after all nine stages
for 18,807 programs at O2 and bit-identical results on 37.4 million specification-function
inputs---}complement these proofs by checking the fidelity of
the migration; correspondence with Rocq remains supported by this
evidence rather than a mechanized equivalence theorem.

We further evaluate whether migration makes the compiler easier for an
agent to improve. We give GPT-6-astra an optimization campaign in each
environment, using the first 24 active hours in both environments. Relative to original CompCert~3.17, the selected Lean
candidate has 11.7\% lower generated-program runtime across 29 small-tests
programs, compared with 2.3\% lower for the Rocq candidate. On ten SPEC
inputs, the corresponding reductions are 6.9\% and 2.2\%. These percentages
follow the rounded ratios in Table~\ref{tab:model-optimization-summary};
they include differences already present in the starting compilers and do
not measure gains relative to each campaign's own baseline. The SPEC measurements
use selected test/train inputs and are not reportable SPEC scores.

The Lean improvements include bounded loop duplication, extended
inlining, instruction scheduling, and revised allocation heuristics,
while preserving the end-to-end correctness theorem and its assumptions.
Both selected candidates pass proof checking, correctness tests, and
the required compilation-cost checks. The optimization traces also show
that the agent produces 106 candidate variants completing full compiler
and proof builds in Lean, compared with 40 in Rocq, under broadly
comparable recorded token usage. This higher throughput gives the agent
more opportunities to evaluate and refine optimizations. These results
demonstrate a practical benefit of the migrated infrastructure in the
observed campaigns, although one campaign per environment cannot
separate the effects of model capability, proof automation, repository
structure, and search choices.

\textbf{Contributions.}
This paper makes the following contributions:

\begin{itemize}
  \item \textbf{A verified C compiler in Lean.} To our knowledge,
  CompCert4Lean is the first compiler implemented and verified in Lean
  that covers the full typed CompCert C abstract syntax and its
  compilation to x86-64 assembly. It supports all 21 configured
  compiler-pass slots and discharges all 4,908 recorded proof
  obligations without project-specific axioms. Combined with the
  CompCert frontend and the system assembler and linker, it produces
  executable Linux x86-64 binaries (\S\ref{sec:implementation}).
  \chen{Differential testing against CompCert~3.17 yields identical RTL
  after all nine stages for every compared program (18,807 at O2 and
  18,808 at O0) and bit-identical results on 37.4 million
  specification-function inputs (\S\ref{sec:eval-differential}).}

  \item \textbf{A parallel workflow for migrating verified developments
  to Lean.} The workflow combines dependency-based parallelism with
  semantic review before large-scale proof reconstruction, reducing
  the risk that translation errors propagate across the development.
  One expert directed parallel agents to complete the main migration in
  approximately six weeks; only three of 1,071 human interactions supplied
  technical direction (\S\ref{sec:human-effort});
  Table~\ref{tab:development-usage} reports the migration usage.

  \item \textbf{Evidence that agents optimize the Lean compiler more
  effectively.} Under matched budgets of 24 active hours, GPT-6-astra produces
  candidates with \textbf{11.7\%} lower generated-program runtime on
  CompCert4Lean and \textbf{2.3\%} lower on \chen{Rocq} CompCert across
  29 small-tests programs, both relative to original CompCert~3.17, while
  preserving the end-to-end correctness theorem in both cases
  (Figure~\ref{fig:intro:teaser}, \S\ref{sec:evaluation}).
\end{itemize}
\section{Overview}
\label{sec:overview}

CompCert is an optimizing C compiler developed and verified in
Rocq~\cite{leroy2009formal,leroy2009backend}. Its verified pipeline consists
of passes over a sequence of intermediate representations (IRs), starting
from the typed CompCert C abstract syntax tree (AST) and ending at a target
assembly AST. Early passes simplify C expressions and control structures
through IRs such as Clight and Cminor. Later passes lower the program to
register-transfer form (RTL), perform optimizations and register
allocation, and generate assembly for the selected backend. Some passes
transform programs within the same IR; others translate between IR levels.

Each IR has an operational semantics defining how its programs execute and
what behaviors they can exhibit. Each compiler pass comes with a
correctness proof relating the input and output programs through a
simulation relation. These proofs allow internal steps and program
representations to change while preserving the required relationship
between observable behaviors. The pass-level results compose into an
end-to-end compiler correctness theorem: when compilation succeeds, the
generated assembly preserves the source program's semantics, subject to
CompCert's treatment of undefined behavior.

The IR semantics and pass-level proofs depend on shared foundations,
including machine arithmetic~\cite{boldo2011flocq}, global environments,
and a common memory model. These foundations connect executable operations
to the invariants used throughout the correctness proofs. Migrating such a
compiler therefore requires preserving not only its IRs and
transformations, but also their semantics, correctness statements,
and the shared definitions on which those
properties depend.

\subsection{Migration Scope and Correctness Obligation}
\label{sec:overview-scope}

We migrate CompCert~3.17's verified compilation pipeline
from the typed CompCert C AST to the x86-64 assembly AST. Our scope
includes the shared foundations, IR semantics, all 21 configured
compiler-pass slots, and the composed semantic-preservation theorem. The
following correctness obligation and verification boundary specify the
result that CompCert4Lean must establish.

\textbf{Correctness obligation.}
Fix the x86-64 configuration and the compiler's explicit interface
parameters and required certificates. Let $\operatorname{compile}$
\chen{denote the migrated pipeline.} We require:
\begin{equation}
  \label{eq:compiler-correctness}
  \begin{gathered}
    \forall\, p:\mathtt{Csyntax.program},\ a:\mathtt{Asm.program},\ \xi,\\
    \operatorname{compile}(p)=\mathrm{OK}(a)
    \;\Longrightarrow\;
    \operatorname{BackwardSim}
    \bigl(\mathcal{C}_{\xi}(p),\,\mathcal{A}_{\xi}(a)\bigr).
  \end{gathered}
\end{equation}
Here $\mathcal{C}_{\xi}$ and $\mathcal{A}_{\xi}$ are the source C and target
assembly operational semantics under the same external-call interface
$\xi$. Following CompCert, backward simulation ensures that every
observable target behavior matches a source behavior, allowing the target
to continue with more defined behavior after the source encounters
undefined behavior. Consequently, if the source program cannot go wrong,
every target behavior is permitted by the source semantics.

\textbf{Verification boundary and assumptions.}
CompCert4Lean adopts CompCert~3.17's verification boundary for the selected
x86-64 pipeline, from the typed CompCert C AST to the x86-64 assembly AST.
Parsing and elaboration precede this boundary; assembly printing,
assembling, and linking follow it and remain outside the compilation
theorem's scope. Provided that the source and target semantics match
CompCert's under the same external-interface assumptions, proving the
correctness obligation above in Lean without project-specific axioms
establishes the same end-to-end semantic-preservation guarantee. This
guarantee does not require correspondence between the intermediate IRs of
the two developments. Our migration scope is broader: we also preserve the
definitions and correctness statements throughout the selected IR
pipeline and reconstruct their proofs in Lean.
\chen{We assess the fidelity of this migration through independent
source-level audits and extensive differential testing against
CompCert~3.17. The tests find identical RTL after all nine compilation
stages for every compared program (18,807 at O2 and 18,808 at O0) and
bit-identical results for 202 executable
specification functions over 37.4 million inputs
(Section~\ref{sec:eval-differential}).}

\textbf{Trusted computing base.} 
Although the migration replaces the entire proof assistant, 
most of the trusted base stays the same. 
As in CompCert, the source and target semantics, the external-call interface~$\xi$, and the unverified code that the port reuses from CompCert 
on either side of the trust boundary, from the preprocessor to the linker, 
are all in the trusted computing base. 
The register allocator is not trusted but instead uses a verified 
validator to check its output (Section~\ref{sec:approach-integration}). 
The CompCert4Lean trusted base differs in three ways. First, the Lean 
kernel and its standard axioms (\texttt{propext}, 
\texttt{Classical.choice}, and \texttt{Quot.sound}) 
replace the Rocq kernel and the axioms that CompCert 
assumes (proof irrelevance, functional extensionality, 
excluded middle, and indefinite description). 
Second, the Lean compiler, the C compiler that 
compiles its output, and the Lean runtime replace 
extraction and the OCaml compiler and runtime. 
Each \texttt{@[csimp]} replacement, which swaps a 
definition for a faster implementation at compile time, 
is proved equal to that definition 
(Section~\ref{sec:approach-integration}), so it adds no trust 
beyond the Lean compiler. Third, the port adds a 
serialization bridge to the reused OCaml front end 
and printer, which CompCert does not need. The bridge 
is unverified, but its codecs are generated from the 
extracted types, and every compilation in our 
differential testing (Section~\ref{sec:eval-differential}) 
passes through it without error.

\subsection{Migration Workflow}
\label{sec:overview-workflow}

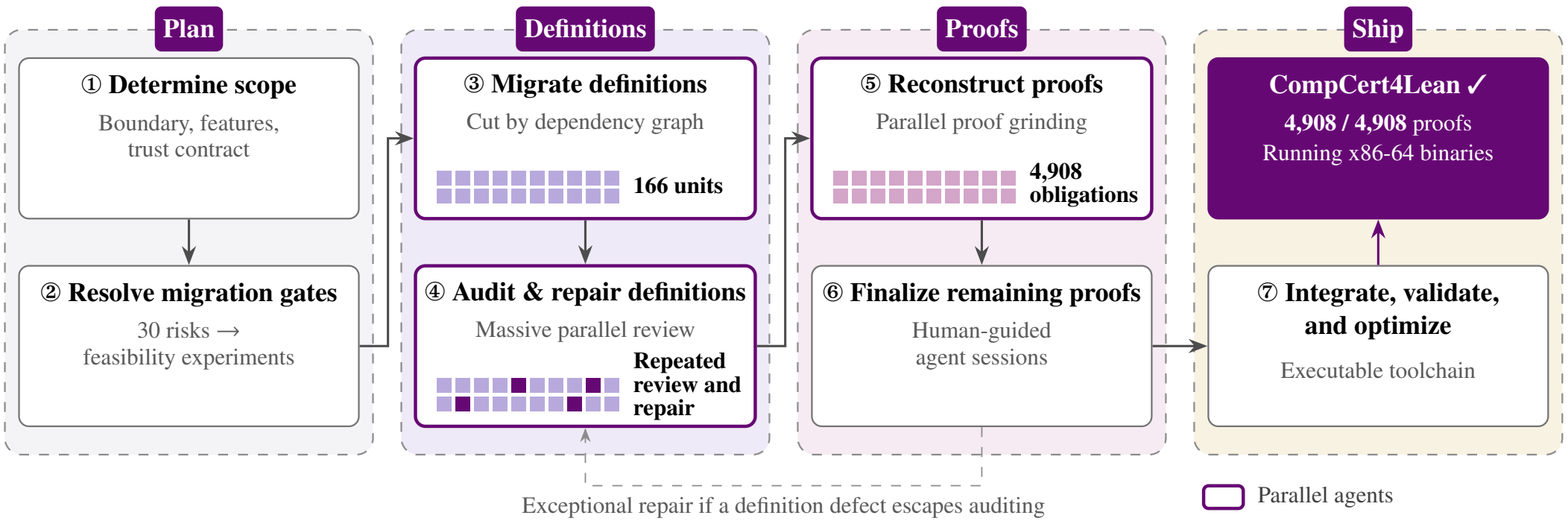
\begin{figure}[t]
  \centering
  \captionsetup{skip=2pt}
  \resizebox{\linewidth}{!}{%
\begin{tikzpicture}[
  font=\figfont,
  stage/.style={draw=black!55, fill=white, rounded corners=3pt,
    line width=0.5pt, minimum width=36mm, minimum height=17mm},
  par/.style={stage, draw=thupurple, line width=1pt},
  title/.style={font=\figfont\bfseries\footnotesize, anchor=north,
    align=center, inner sep=0pt},
  sub/.style={font=\figfont\scriptsize, text=black!65, anchor=north,
    align=center, inner sep=0pt},
  group/.style={draw=black!45, dashed, rounded corners=5pt, fill=#1,
    minimum width=39mm, minimum height=45mm, line width=0.5pt},
  tab/.style={fill=thupurple, text=white, font=\figfont\bfseries\footnotesize,
    rounded corners=2pt, inner sep=2.5pt, text height=1.6ex, text depth=0.4ex},
  flow/.style={-{Stealth[length=2mm]}, line width=0.7pt, draw=black!70},
]
  \foreach \x/\c/\n/\name in {0/figfillN/a/Plan, 4.2/figfillA/b/Definitions,
      8.4/figfillB/c/Proofs, 12.6/figfillC/d/Ship} {
    \node[group=\c] (g\n) at (\x,-1.1) {};
    \node[tab] at (g\n.north) {\name};
  }

  \newcommand{\stagetext}[3]{%
    \node[title] at ([yshift=-5pt]#1.north) {#2};
    \node[sub] at ([yshift=-17pt]#1.north) {#3};}

  \node[stage] (s1) at (0,0) {};
  \stagetext{s1}{\ding{172} Determine scope}{Boundary, features,\\trust contract}
  \node[stage] (s2) at (0,-2.2) {};
  \stagetext{s2}{\ding{173} Resolve migration gates}{30 risks $\rightarrow$\\feasibility experiments}

  \node[par] (s3) at (4.2,0) {};
  \stagetext{s3}{\ding{174} Migrate definitions}{Cut by dependency graph}
  \node[par] (s4) at (4.2,-2.2) {};
  \stagetext{s4}{\ding{175} Audit \& repair definitions}{Massive parallel review}

  \node[par] (s5) at (8.4,0) {};
  \stagetext{s5}{\ding{176} Reconstruct proofs}{Parallel proof grinding}
  \node[stage] (s6) at (8.4,-2.2) {};
  \stagetext{s6}{\ding{177} Finalize remaining proofs}{Human-guided\\agent sessions}

  \node[stage, fill=thupurple, draw=thupurple] (out) at (12.6,0) {};
  \node[title, text=white] at ([yshift=-5pt]out.north) {CompCert4Lean \ding{51}};
  \node[sub, text=white, font=\figfont\scriptsize] at ([yshift=-17pt]out.north)
    {\textbf{4,908 / 4,908} proofs\\[1pt]Running x86-64 binaries};
  \node[stage] (s7) at (12.6,-2.2) {};
  \stagetext{s7}{\ding{178} Integrate, validate,\\and optimize}{}
  \node[sub] at ([yshift=-29pt]s7.north) {Executable toolchain};

  \newcommand{\workgrid}[4]{%
    \foreach \i in {0,...,9} \foreach \j in {0,1} {
      \fill[#2] ([xshift=7pt+\i*5.6pt, yshift=5pt+\j*5.6pt]#1.south west)
        rectangle ++(4.4pt,4.4pt);
    }
    \foreach \i/\j in {#4} {
      \fill[thupurple] ([xshift=7pt+\i*5.6pt, yshift=5pt+\j*5.6pt]#1.south west)
        rectangle ++(4.4pt,4.4pt);
    }
    \node[font=\figfont\bfseries\scriptsize, anchor=west, align=left,
      inner sep=0pt] at ([xshift=66pt, yshift=10.6pt]#1.south west) {#3};}

  \workgrid{s3}{figgridA}{166 units}{}
  \workgrid{s4}{figgridA}{}{1/0,4/1,7/0,8/1}
  \node[font=\figfont\bfseries\scriptsize, anchor=west, align=left,
    inner sep=0pt] at ([xshift=66pt, yshift=12.6pt]s4.south west)
    {Repeated\\[-1pt]review and\\[-1pt]repair};
  \workgrid{s5}{figgridB}{4,908\\[-1pt]obligations}{}

  \draw[flow] (s1) -- (s2);
  \draw[flow] (s2.east) -- ++(0.3,0) |- (s3.west);
  \draw[flow] (s3) -- (s4);
  \draw[flow] (s4.east) -- ++(0.3,0) |- (s5.west);
  \draw[flow] (s5) -- (s6);
  \draw[flow] (s6.east) -- (s7.west);
  \draw[flow, draw=thupurple] (s7) -- (out);

  \draw[-{Stealth[length=1.5mm]}, line width=0.4pt,
    dashed, draw=black!45]
    (s6.south) -- ++(0,-0.62) -| node[pos=0.25, below,
      font=\figfont\scriptsize, text=black!65]
      {Exceptional repair if a definition defect escapes auditing}
    (s4.south);

  \draw[draw=thupurple, line width=1pt, rounded corners=1.5pt, fill=white]
    (10.75,-3.92) rectangle ++(0.42,0.24);
  \node[font=\figfont\scriptsize, anchor=west, text=black!75]
    at (11.2,-3.8) {Parallel agents};
\end{tikzpicture}
}
  \vspace{-4pt}
  \caption{CompCert4Lean migration workflow. \ding{172}--\ding{173}
  Fix scope and resolve risky design choices. \ding{174} Partition the
  dependency graph into 166 units for independent, parallel definition migration.
  \ding{175} Review and repair: 11 completed broad definition-review rounds;
  the first found 1,091 erased theorem statements. \ding{176} Rebuild proofs
  in parallel against frozen statements. \ding{177} Human-guided sessions
  close the remainder. \ding{178} Deliver 4,908/4,908 proved obligations and a
  working toolchain. Purple outlines denote parallel-agent stages; the thin
  dashed arrow denotes exceptional repair of definition defects missed by
  auditing and exposed during proof work.}

  \Description{Seven stages in four phase groups: determine scope;
  resolve migration gates; migrate definitions; audit and repair definitions;
  reconstruct proofs; finalize remaining proofs; and integrate, validate,
  and optimize. Definition and proof reconstruction connect to groups of
  parallel agents. A thin, muted dashed arrow returns from proof finalization
  to definition auditing and repair only if a definition defect escapes
  early auditing.}
  \label{fig:migration-workflow}
\end{figure}

Figure~\ref{fig:migration-workflow} presents our workflow for migrating
CompCert to Lean. It consists of seven stages, grouped into planning,
definition migration, proof reconstruction, and executable integration.
The workflow enables large-scale parallel work by LLM agents
across definition migration, auditing, and proof reconstruction. We outline
these stages below.

\begin{itemize}
  \item \textbf{Resolve migration gates.} We investigate representation and
  interface choices whose consequences could affect many downstream tasks.
  Risk discovery identified 30 issues, which informed feasibility experiments
  covering shared foundations, compiler services, and a backend slice.
  Small executable experiments and checked proofs test these choices before
  broad translation begins, reducing the risk of costly rework across
  parallel agents (Section~\ref{sec:approach-gates}).

  \item \textbf{Migrate definitions.} We partition the development along its
  dependency graph into 166 units: 154 source components and 12 shared
  providers. Agents translate definitions, semantics, and exact theorem
  statements in parallel against stable shared interfaces. They complete
  necessary supporting proofs but defer most proof obligations, allowing
  downstream translation to proceed before upstream proofs are complete
  (Section~\ref{sec:approach-definitions}).

  \item \textbf{Audit \& repair definitions.} Independent parallel agents
  compare the translated definitions and statements with the Rocq source
  before a large proof effort depends on them. The first audit found 1,091
  statements reduced to \texttt{True}, demonstrating why successful builds
  were insufficient. Repeated review and repair included 11 completed broad
  definition-review rounds to identify and correct such errors before proof
  reconstruction (Section~\ref{sec:approach-audit}).

  \item \textbf{Reconstruct proofs.} Automated parallel agents reconstruct
  proofs against the accepted definitions and statements, working from an
  inventory of 4,908 obligations. Bounded attempts and checks of partial
  results allow the process to retain completed proofs while directing
  further effort toward unresolved targets. This stage handles proof
  reconstruction at scale and identifies obligations that require
  human-guided investigation (Section~\ref{sec:approach-proofs}).

  \item \textbf{Finalize remaining proofs.} Human-guided agent sessions
  investigate difficult obligations and develop supporting lemmas to unblock
  dependent proofs, bringing the completed inventory to all 4,908
  obligations. If proof work exposes a definition or statement defect, a
  local repair requires rechecking the affected proofs. This is the
  exceptional fallback shown in Figure~\ref{fig:migration-workflow}, which
  early auditing aims to minimize (Section~\ref{sec:approach-finalize}).

  \item \textbf{Integrate, validate, and optimize.} We check the completed
  proofs and their assumptions, then connect the verified core to an
  executable toolchain that produces running Linux x86-64 binaries.
  End-to-end tests also exercise components beyond the compilation theorem's
  boundary, including parsing, assembly printing, and linking. We improve
  compiler execution and generated-program performance while preserving
  the correctness guarantee (Section~\ref{sec:approach-integration}).
\end{itemize}
\section{Approach}
\label{sec:approach}

We now describe the technical decisions and coordination methods
underlying the workflow in Figure~\ref{fig:migration-workflow}. We focus
on how to establish shared interfaces, audit translated definitions and
statements, and reconstruct proofs in parallel, using concrete migration
challenges to explain these choices.

\subsection{Identify and Resolve Migration Gates}
\label{sec:approach-gates}

Before broad parallel migration, we investigate shared choices that could
invalidate downstream translations and proofs if resolved incorrectly.
An error in a common representation or interface can spread across agents,
making a later correction require widespread rework. Six rounds of risk
discovery identified 30 issues, informing five feasibility gates:
representations (Gate~1), floating point (Gate~2), memory and injections
(Gate~3), validated compiler services and successful execution (Gate~4),
and a backend slice through pass interfaces and simulation composition
(Gate~5). Executable experiments and checked proofs test these choices
before agents build on them. The agents autonomously proposed and
implemented the technical solutions to all identified migration gates.
Appendix~\ref{app:investigation-gates} describes the floating-point and
memory/injection gates and reports the effort directly attributed to
Gates~1--3.

These investigations illustrate four recurring sources of difficulty in
Rocq-to-Lean migration, which can arise together within a gate.
\emph{Proof-assistant expressiveness gaps} require alternative encodings:
CompCert's infinite event traces are coinductive data, whereas Lean~4.33's
coinductive mechanism supports predicates in \texttt{Prop}, not data in
\texttt{Type}. \emph{Library semantic mismatches} require checking observable
behavior, including NaN signs and payloads and floating-point rounding,
rather than substituting operations with similar names.
\emph{Missing proof infrastructure} can block downstream proofs even when
the translated operations have the intended behavior: supporting library
lemmas must also be reconstructed. Finally, \emph{representation and
invariant mismatches} arise when a replacement data structure reproduces
individual operations but does not support their required relationships;
CompCert's memory proofs depend on permissions, non-overlap, and
preservation of memory injections. These sources of difficulty guide what
a feasibility experiment must establish, beyond whether its code compiles.

\textbf{Example: coinductive traces and infinite behavior.}
The representation gate illustrates how an expressiveness gap becomes a
question about downstream semantic proofs. CompCert uses coinduction both
for infinite executions, which are propositions, and for infinite event
traces, which are data appearing in program behaviors. Lean~4.33's support
for coinductive predicates therefore did not by itself provide a
representation for those traces. The gate had to test an alternative
encoding together with the semantic arguments that consume it.

\begin{wrapfigure}[12]{r}{0.4\linewidth}
\begin{lstlisting}[style=leanexample,aboveskip=0pt,belowskip=0pt]
abbrev TraceInf (Event : Type) :=
  Nat -> Event
def head (stream : TraceInf Event) :
    Event := stream 0
def tail (stream : TraceInf Event) :
    TraceInf Event :=
  fun index => stream (index + 1)
def cons (event : Event)
    (stream : TraceInf Event) :
    TraceInf Event
  | 0 => event
  | index + 1 => stream index
\end{lstlisting}
\end{wrapfigure}
We represented an infinite trace as a total function mapping each
natural-number position to its event.
All observations are total. CompCert also corecursively constructs streams of
nonempty finite trace chunks. We represented these as functions from
natural numbers to nonempty chunks and flattened them with a cursor that
recurses on the requested event position. Nonemptiness ensures that each
recursive call consumes an event, allowing the flattened stream to supply
any requested observation.

However, an encoding of streams alone would not establish that the
behavior proofs could be reconstructed. The checked prototype established
finite-prefix and flattening laws, then reconstructed the behavior-existence
argument: repeatedly choose a reachable finite execution with a nonempty
trace, construct the resulting chunk stream, and prove coinductively that
its flattening is the trace of an infinite reactive execution.
Infinite-execution predicates remained coinductive in \texttt{Prop}.
This argument connected the alternative data representation to the
semantic reasoning required by its consumers.

The encoding also exposed a subtler issue: equality. Function
extensionality identifies streams with identical observations, whereas
Rocq's equality on coinductive data does not automatically identify
bisimilar streams. We retained the source's named bisimulation relations,
proved their correspondence to pointwise observation equality in Lean,
and reviewed consumers that used ordinary trace or behavior equality.
These checks made the representation change explicit rather than silently
assuming that the two notions of equality were interchangeable.

Resolving both the construction and equality questions before broad
migration gave agents a shared interface supported by checked semantic
proofs, reducing the risk of incompatible downstream encodings. The gate
established feasibility for the checked interface and consumers; it did
not establish equivalence between the complete Rocq and Lean developments.
The later production translation remained subject to its own correspondence
audit (Section~\ref{sec:approach-audit}).

\subsection{Migrate Definitions in Parallel}
\label{sec:approach-definitions}

The workflow next migrates definitions in parallel using the translation
loop in Algorithm~\ref{alg:definition-migration} (lines~1--13). We partition
the dependency graph into 166 units---154 source components and 12 shared
providers---with one canonical owner per declaration and provider, so
workers share stable interfaces. Lines~1--2 initialize the baseline and
pending units; lines~4--5 launch units whose prerequisites are integrated
and whose dependencies have no unresolved blockers. Each assignment uses
an isolated checkout of a pinned baseline and reserves disjoint writable
files. Agents translate executable definitions, semantic relations, and
exact theorem statements, recording most proofs as outstanding obligations
while completing supporting proofs required by interfaces. This separation
allows downstream translation to proceed before all upstream proofs are
complete, without allowing consumers to bypass required proof evidence.

\begin{algorithm}[t]
  \caption{Parallel definition migration and iterative correspondence auditing.}
  \label{alg:definition-migration}
  \small
  \setlength{\parindent}{0pt}
  \setlength{\tabcolsep}{0pt}
  \noindent\begin{tabular}{@{}p{4.5em}p{\dimexpr\linewidth-4.5em\relax}@{}}
    \textbf{Input:} & pinned source $\mathit{source}$, contract $\mathit{contract}$,
      dependency graph $\mathit{graph}$. \\[2pt]
    \textbf{Output:} & accepted baseline, obligation inventory, audit evidence
      and decision; or a blocking report.
  \end{tabular}\par
  \vspace{3pt}
  \begin{tabbing}
    \hspace{2.4em}\=\hspace{1.5em}\=\hspace{1.5em}\=\hspace{1.5em}\=\kill
    \makebox[1.6em][r]{1}\> $\mathit{base} \gets \textsc{InitialBaseline}(\mathit{contract})$ \\
    \makebox[1.6em][r]{2}\> $\mathit{pending} \gets \textsc{Units}(\mathit{graph});\quad \mathit{running} \gets \emptyset$ \\
    \makebox[1.6em][r]{3}\> \textbf{while} $\mathit{pending} \ne \emptyset$: \\
    \makebox[1.6em][r]{4}\>\> $\mathit{ready} \gets \textsc{ReadyUnits}(\mathit{pending},\mathit{graph},\mathit{base})$ \\
    \makebox[1.6em][r]{5}\>\> $\mathit{running} \gets \textsc{LaunchLeased}(\mathit{ready},\mathit{base},\mathit{running})$ \\
    \makebox[1.6em][r]{6}\>\> \textbf{if} $\mathit{running}=\emptyset$: \textbf{return} $\textsc{Blocked}(\mathit{base},\mathit{pending})$ \\
    \makebox[1.6em][r]{7}\>\> $\mathit{results} \gets \textsc{CollectFinished}(\mathit{running})$ \\
    \makebox[1.6em][r]{8}\>\> \textbf{for each} $\mathit{candidate} \in \mathit{results}$: \\
    \makebox[1.6em][r]{9}\>\>\> \textbf{if} $\textsc{IntegrationReady}(\mathit{candidate},\mathit{base},\mathit{contract})$: \\
    \makebox[1.6em][r]{10}\>\>\>\> $\mathit{base} \gets \textsc{Integrate}(\mathit{base},\mathit{candidate})$ \\
    \makebox[1.6em][r]{11}\>\>\>\> $\mathit{pending} \gets \mathit{pending} \setminus \textsc{Units}(\mathit{candidate})$ \\
    \makebox[1.6em][r]{12}\>\>\> \textbf{else}: $\textsc{RecordRetryOrBlock}(\mathit{candidate},\mathit{pending})$ \\
    \makebox[1.6em][r]{13}\>\> $\textsc{ReleaseFiles}(\mathit{results},\mathit{running})$ \\[3pt]
    \makebox[1.6em][r]{14}\> $\mathit{evidence} \gets \emptyset;\quad \mathit{scope} \gets \textsc{WholeDevelopment}$ \\
    \makebox[1.6em][r]{15}\> \textbf{loop}: \\
    \makebox[1.6em][r]{16}\>\> $\mathit{snapshot} \gets \textsc{Snapshot}(\mathit{base})$ \\
    \makebox[1.6em][r]{17}\>\> $\mathit{inventory} \gets \textsc{ExtractObligations}(\mathit{snapshot},\mathit{source})$ \\
    \makebox[1.6em][r]{18}\>\> $\mathit{evidence} \gets \textsc{AuditRound}(\mathit{snapshot},\mathit{source},$ \\
    \>\>\> $\mathit{scope},\mathit{inventory},\mathit{evidence})$ \\
    \makebox[1.6em][r]{19}\>\> $\mathit{decision} \gets \textsc{ReviewFindings}(\mathit{evidence},\mathit{contract})$ \\
    \makebox[1.6em][r]{20}\>\> \textbf{if} $\mathit{decision.unresolvedCriticalIssues}=\emptyset$: \\
    \makebox[1.6em][r]{21}\>\>\> \textbf{return} $(\mathit{snapshot},\mathit{inventory},\mathit{evidence},\mathit{decision})$ \\
    \makebox[1.6em][r]{22}\>\> \textbf{if} $\mathit{decision.needsOwner}$: \textbf{return} $\textsc{AwaitDecision}(\mathit{snapshot},\mathit{evidence})$ \\
    \makebox[1.6em][r]{23}\>\> $\mathit{edits} \gets \textsc{ParallelRepair}(\mathit{snapshot},\mathit{decision.repairs})$ \\
    \makebox[1.6em][r]{24}\>\> $\mathit{base} \gets \textsc{IntegrateCheckedRepairs}(\mathit{base},\mathit{edits})$ \\
    \makebox[1.6em][r]{25}\>\> $\mathit{changed} \gets \textsc{ChangedInputs}(\mathit{snapshot},\mathit{base})$ \\
    \makebox[1.6em][r]{26}\>\> $\mathit{evidence} \gets \textsc{InvalidateAffected}(\mathit{evidence},\mathit{changed})$ \\
    \makebox[1.6em][r]{27}\>\> $\mathit{scope} \gets \textsc{NextAuditScope}(\mathit{changed},\mathit{evidence})$
  \end{tabbing}
\end{algorithm}

As workers finish, lines~7--13 check and integrate their results, making
accepted providers available to dependent units. The integration check
(line~9) enforces canonical ownership, intended declaration identities and
statement interfaces, restrictions on proof escapes, and build requirements
against the current baseline, including changes accepted since the worker
started. Rejected candidates retain diagnostics for retry or a blocking
report, while unrelated ready work continues. These checks establish that
translations can be assembled and used together, but do not establish
source correspondence. The audit loop (lines~14--27), described in
Section~\ref{sec:approach-audit}, determines whether the definitions and
statements can be accepted as the baseline for proof reconstruction.

\subsection{Iteratively Audit and Repair Definitions}
\label{sec:approach-audit}

The workflow next audits and repairs the translated definitions and
statements using Algorithm~\ref{alg:definition-migration}'s second loop.
Lines~14--18 initialize the audit evidence, fix a snapshot, and extract its
proof inventory for independent parallel review. Reviewers compare
computational definitions, theorem statements, and relevant consumer
interfaces with the Rocq source, including premises, binders, and
proof-bearing invariants. At line~19, findings are classified by severity
and unresolved critical issues from all rounds are retained in
$\mathit{decision.unresolvedCriticalIssues}$. In our migration, the human did not escalate audit issues. Human
involvement included only three technical-direction messages, including
one confirmed override, as detailed in Section~\ref{sec:human-effort}. The stopping condition at line~20
requires that no unresolved critical issues remain and freezes the reviewed
targets for ordinary proof reconstruction; it does not imply complete
review coverage or discharge the remaining proofs.

Otherwise, questions requiring an expert decision are returned at line~22,
and repairs proceed in parallel against a common snapshot with disjoint
writable files (lines~23--24). After integration, lines~25--27 invalidate
evidence affected by changed definitions or dependencies and select the
next review scope. A provider change can invalidate a consumer's review
even when the consumer itself is unchanged. Rechecking affected interfaces
and reusing only unchanged evidence allows review and repair to iterate
without treating an earlier successful review as permanent acceptance.
Appendix~\ref{app:audit-rounds} reports how findings evolved across
review rounds and records the evidence and limitations of the final
acceptance decision.

\phantomsection\label{ex:constructor-interface}
\textbf{Example: a lost constructor interface.}
During our migration campaign, the third proof-inventory review found that \texttt{SetoidList.InA} had
been translated directly as
$\exists y,\; \mathit{eqA}\;x\;y \land y \in \mathit{xs}$.
The Rocq source instead defines the relation inductively with head and
tail constructors, then proves \texttt{InA\_alt} to characterize it by
this existential predicate. In the translation, \texttt{InA\_alt} followed
immediately by \texttt{Iff.rfl}, but the constructor and induction
interfaces were lost. The predicates were logically equivalent; the
mismatch concerned the interface available to downstream proofs.

The agents restored the inductive definition and constructors, proved the
characterization by induction, and adjusted membership and sortedness
proofs in the consuming \texttt{MSetGenTree} unit. Iterative review caught a lost proof interface despite logical equivalence
and successful builds; the repair restored the constructors and induction
interface needed by downstream proofs.

\subsection{Reconstruct Proofs in Parallel}
\label{sec:approach-proofs}

Algorithm~\ref{alg:proof-grinding} reconstructs proofs in parallel against
the accepted definitions and statements. Lines~1--6 initialize the proof
state and assign ready targets whose dependencies are available, whose
attempt budgets remain, and which are not already assigned. Workers use
isolated baseline revisions and disjoint writable files, adapting source
proofs to Lean's libraries and automation while preserving the frozen
targets. Attempts are bounded so that difficult obligations do not prevent
progress on other available proofs. Operational failures enter recovery
(lines~10--11), while suspected definition mismatches return to review
(line~13); neither justifies weakening the target statement.

\begin{algorithm}[t]
  \caption{Automated proof grinding with partial acceptance and bounded retries.}
  \label{alg:proof-grinding}
  \small
  \setlength{\parindent}{0pt}
  \setlength{\tabcolsep}{0pt}
  \noindent\begin{tabular}{@{}p{4.5em}p{\dimexpr\linewidth-4.5em\relax}@{}}
    \textbf{Input:} & accepted baseline $\mathit{base}$, exact target inventory
      $\mathit{inventory}$, attempt policy $\mathit{policy}$. \\[2pt]
    \textbf{Output:} & extended baseline, completed targets, and explicit
      unfinished targets.
  \end{tabular}\par
  \vspace{3pt}
  \begin{tabbing}
    \hspace{2.4em}\=\hspace{1.5em}\=\hspace{1.5em}\=\hspace{1.5em}\=\hspace{1.5em}\=\kill
    \makebox[1.6em][r]{1}\> $\mathit{state} \gets \textsc{InitProofState}(\mathit{base},\mathit{inventory},\mathit{policy})$ \\
    \makebox[1.6em][r]{2}\> $\mathit{running} \gets \emptyset$ \\
    \makebox[1.6em][r]{3}\> \textbf{loop}: \\
    \makebox[1.6em][r]{4}\>\> $\mathit{ready} \gets \textsc{ReadyTargets}(\mathit{state})$ \\
    \makebox[1.6em][r]{5}\>\> $\mathit{batches} \gets \textsc{GroupByOwnerAndFile}(\mathit{ready},\mathit{policy})$ \\
    \makebox[1.6em][r]{6}\>\> $\mathit{running} \gets \textsc{LaunchLeased}(\mathit{batches},\mathit{base},\mathit{running})$ \\
    \makebox[1.6em][r]{7}\>\> \textbf{if} $\mathit{running}=\emptyset$: \\
    \makebox[1.6em][r]{8}\>\>\> \textbf{return} $(\mathit{base},\textsc{Finished}(\mathit{state}),\textsc{Unfinished}(\mathit{state}))$ \\
    \makebox[1.6em][r]{9}\>\> $\mathit{result} \gets \textsc{CollectNext}(\mathit{running})$ \\
    \makebox[1.6em][r]{10}\>\> \textbf{if} $\textsc{OperationalFailure}(\mathit{result})$: \\
    \makebox[1.6em][r]{11}\>\>\> $\textsc{RecordIncidentAndRecover}(\mathit{result},\mathit{state})$ \\
    \makebox[1.6em][r]{12}\>\> \textbf{else}: \\
    \makebox[1.6em][r]{13}\>\>\> $\textsc{RecordDefinitionFindings}(\mathit{result},\mathit{state})$ \\
    \makebox[1.6em][r]{14}\>\>\> $\mathit{accepted} \gets \emptyset$ \\
    \makebox[1.6em][r]{15}\>\>\> $\mathit{extensions} \gets \textsc{ProofCheckpoints}(\mathit{result})$ \\
    \makebox[1.6em][r]{16}\>\>\> \textbf{for each} $\mathit{extension} \in \textsc{DependencyOrder}(\mathit{extensions})$: \\
    \makebox[1.6em][r]{17}\>\>\>\> $\mathit{staged} \gets \textsc{Apply}(\mathit{base},\mathit{extension})$ \\
    \makebox[1.6em][r]{18}\>\>\>\> \textbf{if} $\textsc{ValidExtension}(\mathit{staged},\mathit{base},\mathit{inventory})$: \\
    \makebox[1.6em][r]{19}\>\>\>\>\> $\mathit{base} \gets \mathit{staged}$ \\
    \makebox[1.6em][r]{20}\>\>\>\>\> $\mathit{accepted} \gets \mathit{accepted} \cup \textsc{Targets}(\mathit{extension})$ \\
    \makebox[1.6em][r]{21}\>\>\>\> \textbf{else}: $\textsc{RetainRejectedCandidate}(\mathit{extension})$ \\
    \makebox[1.6em][r]{22}\>\>\> $\textsc{MarkFinished}(\mathit{state},\mathit{accepted})$ \\
    \makebox[1.6em][r]{23}\>\>\> $\mathit{remaining} \gets \textsc{Assigned}(\mathit{result}) \setminus \mathit{accepted}$ \\
    \makebox[1.6em][r]{24}\>\>\> $\textsc{ScheduleSingletonOrDefer}(\mathit{state},\mathit{remaining},\mathit{result})$ \\
    \makebox[1.6em][r]{25}\>\> $\textsc{ReleaseFiles}(\mathit{result},\mathit{running})$ \\
    \makebox[1.6em][r]{26}\>\> $\textsc{Checkpoint}(\mathit{base},\mathit{state})$
  \end{tabbing}
\end{algorithm}

Lines~15--22 retain checked partial progress: a worker may contribute
completed proofs and their required helpers even when other assigned
targets remain open. Each addition is checked in dependency order against
the current integrated baseline, including proofs accepted while the
worker was running. The acceptance check at line~18 requires exact target
types, preservation of existing definitions and proofs, a successful
build, and only the permitted transitive axioms: \texttt{propext},
\texttt{Classical.choice}, and \texttt{Quot.sound}. Unfinished obligations
remain available for further attempts or are deferred when dependencies
or budgets prevent progress (lines~23--26); the algorithm returns both
completed and unfinished targets when no worker remains (lines~7--8).
The human expert can also choose when to stop automatic reconstruction.
In our campaign, the expert stopped it with work still queued and 471
obligations remaining, motivating guided investigation of difficult or
blocked targets (Section~\ref{sec:approach-finalize}).
Appendix~\ref{app:proof-progress} reports the proof-inventory milestones
and the distribution of the remaining obligations.

Batch size balances shared context against the context window available
to each agent. At line~5, we group up to four related proof obligations
from the same file, allowing a worker to reuse definitions, lemmas, and
proof strategies. Larger batches risk exhausting the context window
before all targets are addressed; smaller batches repeat shared context
and waste agent effort. Four was our practical choice for this trade-off,
not an experimentally established optimum. Targets left unfinished by a
batch receive subsequent singleton attempts (line~24), allowing focused
work without repeatedly carrying the entire batch's context.

\subsection{Finalize Proofs through Human-Guided Agent Sessions}
\label{sec:approach-finalize}

The 471 obligations remaining when automatic reconstruction stopped became
inputs to human-guided agent sessions. Each session focused on one file,
with independent sessions able to proceed in parallel. In this migration,
the expert managed context by starting fresh sessions when contexts grew
long, decided whether to batch or continue work, permitted longer timeouts,
and prompted agents to finish proofs in specific files. The agents supplied
proof strategies and constructed proofs and supporting lemmas; the expert
did not contribute technical proof solutions. This process completed all 4,908 recorded
obligations. Section~\ref{sec:human-effort} analyzes the human interactions
throughout the migration. During this stage, only two floating-point-related cases
required local repair: a conversion omitted destination normalization,
and a theorem omitted its format assumptions.
The missing-assumptions case is detailed below;
Appendix~\ref{app:float-repairs} gives the conversion counterexample and repair.

\textbf{A theorem missing its format assumptions.}
A separate failure concerned the Binary theorem that a canonical value
below $2^{\mathit{emax}}$ satisfies the format's boundedness test. Its
translated statement omitted the source assumptions
$0 < \mathit{prec}$ and $\mathit{prec} < \mathit{emax}$. Section-local
Lean instances were not retained as parameters because the declaration
did not explicitly refer to them. Agents constructed a kernel-checked
counterexample with precision 2, maximum exponent 1, mantissa 1, and
exponent 0. The value is canonical and satisfies $1 < 2$, but the
boundedness test requires $0 \leq -1$ and returns false. This configuration
violates the source format assumptions, so it refutes the translated
statement rather than the upstream theorem.

This example demonstrates our workflow's ability to distinguish an
incorrectly translated proof target from a difficult proof obligation:
the agents constructed a kernel-checked counterexample that justified
repairing the statement before resuming proof reconstruction.
The repair explicitly bound both assumptions in the statement and
reused the corresponding SingleNaN theorem to supply its proof. Together with the conversion repair in
Appendix~\ref{app:float-repairs}, this case exercises the feedback path in
Figure~\ref{fig:migration-workflow}: counterexamples justify reopening an
accepted definition or proposition through the review procedure in
Section~\ref{sec:approach-audit}, followed by authorized repair and
rechecking of affected proofs. A stalled search alone would not justify
either change.

\subsection{Integrate, Validate, and Optimize the Result}
\label{sec:approach-integration}

With the correctness obligation in Section~\ref{sec:overview-scope}
discharged, we integrate the verified core into a working C-to-native
toolchain.

\textbf{Integration and validation.}
We assemble the core with the pinned CompCert frontend and assembly
printer, typed interchange formats, concrete compiler services, and the
system assembler and linker. A Lean-native register allocator supplies
proposals to the existing validator. Together, these components support
compilation from C source to native executables. Because components
outside the formal boundary still require testing, we validate the
integrated toolchain; Section~\ref{sec:eval-release} presents the
validation procedures and results.

\textbf{Register allocation.}
We preserve CompCert 3.17's verification boundary: the allocation algorithm
is unverified, but its output must pass the verified RTL-to-LTL validator.
Correctness therefore depends on successful validation rather than on a
proof of the allocation algorithm itself. Builtin expansion requires
additional care because its concrete instruction sequences impose
operand-placement constraints beyond the abstract builtin semantics.
As in CompCert 3.17, our allocator respects these target-specific
constraints; we also check compatibility before expansion. This executable
check guards the integration with the assembly expander without extending
the formal guarantee beyond the assembly abstract syntax described in
Section~\ref{sec:overview-scope}.

\textbf{Compiler execution-time optimization.}
The earlier stages prioritize faithful formal definitions and correctness
proofs over execution efficiency. The initial implementation compiled some
programs too slowly to complete practical validation runs. Straightforward
implementations chosen for ease of formalization therefore required
substantial execution-time optimization before the compiler was practical
to use. Once the toolchain runs, we task agents
specifically with reducing the compiler's own execution time while
preserving the logical definitions and existing proofs. Where appropriate,
Lean's \texttt{@[csimp]} mechanism replaces executable implementations
using kernel-checked proofs of equality with the logical definitions.
For example, interference-graph construction unnecessarily copied an
entire adjacency array because an update preceded a read from the old
array. Reading both neighbor sets before either update avoids this copy,
yet the revised operation is definitionally equal to the original.
\section{Implementation}
\label{sec:implementation}

\begin{table}[!htbp]
  \centering
  \caption{Lean implementation size and proof coverage at the release
  implementation milestone (\texttt{801facb}). Lines are physical source
  lines, including comments and blank lines. Counts exclude tests, external
  dependencies, generated build artifacts, and OCaml/Python integration code.
  A dash indicates a component outside the original migration proof inventory.}
  \label{tab:implementation-statistics}
  \begin{tabular*}{\linewidth}{@{\extracolsep{\fill}}lrrr@{}}
    \toprule
    Component & Lean files & Lines & \shortstack[r]{Completed\\obligations} \\
    \midrule
    Foundations and libraries & 34 & 64,557 & 1,266 \\
    Floating-point development (Flocq) & 24 & 51,178 & 554 \\
    Common semantics and infrastructure & 18 & 44,081 & 799 \\
    Typed-C frontend and proofs & 17 & 46,383 & 437 \\
    Backend transformations and proofs & 56 & 136,448 & 1,516 \\
    x86-64 backend and configuration & 21 & 47,439 & 324 \\
    Pipeline composition and correctness & 3 & 1,877 & 12 \\
    \midrule
    \textbf{Migrated core subtotal} & \textbf{173} & \textbf{391,963} & \textbf{4,908} \\
    \addlinespace
    Native allocator & 8 & 1,603 & --- \\
    Lean toolchain integration & 7 & 2,403 & --- \\
    Build aggregation root & 1 & 191 & --- \\
    \midrule
    \textbf{Total} & \textbf{189} & \textbf{396,160} & \textbf{4,908} \\
    \bottomrule
  \end{tabular*}
  \vspace{-6pt}%
\end{table}

Using the workflow described in Section~\ref{sec:approach}, we implemented
CompCert4Lean, a Lean port of CompCert~3.17's verified compiler core.
The migration agents ran in Codex, powered by GPT-5.6~Sol before
September~17, 2026, and GPT-6-astra thereafter. Both models used the extra-high reasoning setting
(\texttt{xhigh}) by default.
Throughout the migration, we maintained a declaration map linking the
required Rocq definitions and theorems to their Lean counterparts,
recording their ownership and proof obligations. Coverage followed the
selected compiler features and their dependencies across executable
definitions, theorem statements, and proofs, including shared libraries
and generated declarations. Explicit exclusions and reviewed
representation changes distinguished intentional differences from missing
translations.

Table~\ref{tab:implementation-statistics} summarizes the resulting
implementation. The migrated core spans 173 Lean modules and 391,963
physical source lines, compared with 149,038 source lines in the
corresponding 154 Rocq modules. The Lean count includes comments and blank
lines; the core subtotal excludes the added allocator and toolchain
integration. The core covers foundations and floating-point arithmetic,
language semantics, compiler transformations, and the composed correctness
theorem. It discharges all
4,908 recorded obligations, which include theorem proofs and proof-bearing
certificates but do not count every supporting lemma. Adding the native
register allocator and toolchain integration brings the implementation to
189 Lean files and 396,160 lines. These additional components support a
runnable Linux x86-64 compiler, while the formal correctness boundary
remains typed CompCert C abstract syntax to x86-64 assembly abstract
syntax, as specified in Section~\ref{sec:overview-scope}.

\begin{table}[!htbp]
  \centering
  \caption{Recorded development sessions and LLM token usage through the
  \texttt{rc-0.0.1} release snapshot. Token counts are in millions, rounded
  independently to one decimal place; output includes reasoning tokens.
  Sessions are deduplicated and include recorded retries and unsuccessful
  attempts. Usage coverage is incomplete, as described in the text.}
  \label{tab:development-usage}
  \begin{tabular*}{\linewidth}{@{\extracolsep{\fill}}lrrrrr@{}}
    \toprule
    Agent role & \shortstack[r]{Distinct\\sessions}
      & \shortstack[r]{Uncached\\input} & \shortstack[r]{Cached\\input}
      & Output & Total \\
    \midrule
    Interactive / coordinator & 141 & 316.0 & 16,518.5 & 33.3 & 16,867.8 \\
    Built-in subagents & 55 & 14.9 & 478.8 & 1.9 & 495.5 \\
    External workers & 12,239 & 4,874.0 & 61,037.9 & 253.0 & 66,164.9 \\
    \midrule
    \textbf{Total} & \textbf{12,435} & \textbf{5,204.9}
      & \textbf{78,035.2} & \textbf{288.2} & \textbf{83,528.3} \\
    \bottomrule
  \end{tabular*}
  \vspace{-6pt}%
\end{table}

\textbf{Migration cost.}
Table~\ref{tab:development-usage} reports recorded LLM usage by agent role
across 12,435 distinct project sessions. Recorded input and output total
83.53 billion tokens; cached input accounts for 93.75\% of all input
tokens. The accounting ends at the conclusion of the turn that created and
locally verified the annotated \texttt{rc-0.0.1} tag for the September 24
implementation milestone in Table~\ref{tab:implementation-statistics}.
Thirteen interactive/coordinator sessions and 2,989 external-worker sessions
\chen{lack positive usage counters. The figures describe locally
recorded usage with partial coverage of provider billing.}

External workers account for approximately 79\% of the recorded tokens
in Table~\ref{tab:development-usage}. Most recorded usage therefore went
to worker tasks that could run in parallel, reflecting the workflow's
emphasis on decomposing migration work across agents.
Recorded usage is also dominated by input tokens. This pattern is
consistent with proof migration, where agents consult substantial source
and dependency context to construct comparatively short definitions or
proofs.

For Table~\ref{tab:phase-token-usage}, attribution follows recorded task
objectives and includes supporting work. Activity dates span the earliest
and latest attributed activity markers. Manual user-prompt tokens are
counted with \texttt{o200k\_base}.

\begin{table}[!htbp]
  \centering
  \caption{Recorded LLM token usage by migration phase, using the same release
  snapshot as Table~\ref{tab:development-usage}. LLM token counts are in
  millions, rounded independently; output includes reasoning tokens.
  Manual prompt counts are exact and excluded from Total (M).
  Activity dates are inclusive UTC dates. The usage-coverage limitations
  of Table~\ref{tab:development-usage} also apply.}
  \label{tab:phase-token-usage}
  \begin{tabular*}{\linewidth}{@{\extracolsep{\fill}}llrrrrr@{}}
    \toprule
    Phase & \shortstack[l]{Activity\\dates (UTC)} & \shortstack[r]{Uncached\\input (M)}
      & \shortstack[r]{Cached\\input (M)} & Output (M) & Total (M)
      & \shortstack[r]{Manual prompts\\(exact tokens)} \\
    \midrule
    1 & Aug.~11--17 & 9.9 & 296.4 & 1.6 & 307.9 & 2,199 \\
    2 & Aug.~11--16 & 150.5 & 8,602.8 & 25.8 & 8,779.1 & 1,721 \\
    3 & Aug.~17--20 & 106.8 & 2,041.6 & 7.3 & 2,155.6 & 2,433 \\
    4 & Aug.~20--Sep.~2 & 3,239.0 & 45,582.6 & 165.1 & 48,986.7 & 7,260 \\
    5 & Sep.~2--21 & 1,642.2 & 20,412.0 & 82.4 & 22,136.6 & 12,000 \\
    6 & Sep.~21--22 & 36.8 & 705.0 & 3.5 & 745.4 & 1,279 \\
    7 & Sep.~22--24 & 19.6 & 394.9 & 2.5 & 417.0 & 2,673 \\
    \midrule
    \textbf{Total} & \textbf{All} & \textbf{5,204.9} & \textbf{78,035.2}
      & \textbf{288.2} & \textbf{83,528.3} & \textbf{29,565} \\
    \bottomrule
  \end{tabular*}
  \par\smallskip
  {\footnotesize\raggedright
  \textit{Note:} Phase~4 dates cover its main activity. Minor local
  definition fixes were subsequently made to address issues discovered
  during phase~6.\par}
  \vspace{-6pt}%
\end{table}

Table~\ref{tab:phase-token-usage} attributes the same recorded usage
across the seven workflow phases in Figure~\ref{fig:migration-workflow}
by recorded task objectives.
Review and repair (phase~4) account for approximately 59\% of the recorded
tokens in Table~\ref{tab:phase-token-usage}, with activity spanning
August~20 to September~2. However, the last issue whose
repair had broad downstream consequences was identified in the third
review round; later issues required local fixes. We conservatively
completed eleven broad review rounds (Appendix~\ref{app:audit-rounds}).
\chen{In retrospect, this suggests that stopping earlier could reduce review
cost, although the appropriate stopping point was uncertain}
at the time. Reducing review also risks cascading repairs during
downstream proof reconstruction, when many proofs already depend on the
shared definitions. The cost therefore reflects the effort to establish
stable definitions before large-scale proof work.

\subsection{Human Effort}
\label{sec:human-effort}
The migration involved 29,565 tokens of human-authored prompts across the
six-week development period (Table~\ref{tab:phase-token-usage}), a modest
volume of direct instruction for a development of this scale. One author
served as the sole human developer directing the main migration, with
agents constructing definitions and proofs. Human interaction focused on
setting priorities, reviewing results, and directing subsequent work. In the
expert's experience, parallel workers often ran autonomously for hours
between interventions, with human attention concentrated around progress
checks and phase or review-round boundaries.

Human involvement was therefore intermittent across the six weeks,
rather than continuous implementation work. The workflow allowed the
expert to guide many concurrent tasks while agents constructed definitions
and proofs. Under one expert's direction, agents migrated a mature verified
compiler embodying an estimated several dozen person-years of accumulated
development into Lean in approximately six weeks.

\begin{wrapfigure}[10]{r}{0.43\linewidth}
  \centering
\begin{tikzpicture}[font=\footnotesize]
  \definecolor{roleworkflow}{HTML}{6B91B4}
  \definecolor{rolereview}{HTML}{E4B45A}
  \definecolor{roledirection}{HTML}{8E354A}
  \pgfmathsetmacro{\directionend}{3/1071*360}
  \pgfmathsetmacro{\reviewend}{120/1071*360}
  \fill[roleworkflow] (0,0) -- (\reviewend:0.7cm)
    arc[start angle=\reviewend,end angle=360,radius=0.7cm] -- cycle;
  \begin{scope}[shift={({(\directionend+\reviewend)/2}:0.10cm)}]
    \fill[rolereview] (0,0) -- (\directionend:0.7cm)
      arc[start angle=\directionend,end angle=\reviewend,radius=0.7cm] -- cycle;
  \end{scope}
  \begin{scope}[shift={({\directionend/2}:0.24cm)}]
    \fill[roledirection] (0,0) -- (0.7,0)
      arc[start angle=0,end angle=\directionend,radius=0.7cm] -- cycle;
  \end{scope}
  \draw[black!45,line width=0.25pt] (\reviewend:0.7cm) arc[start angle=\reviewend,end angle=360,radius=0.7cm];
  \draw[black!65,line width=0.4pt] (-0.7,0) -- (-0.85,-0.8) -- (0.85,-0.8);
  \node[anchor=west,align=left] at (0.9,-0.8)
    {Workflow: 951 (88.8\%)};
  \draw[black!65,line width=0.4pt] ({(\directionend+\reviewend)/2}:0.8cm) -- (1.0,0.65) -- (1.05,0.65);
  \node[anchor=west,align=left] at (1.1,0.65)
    {Technical review/discussion\\117 (10.9\%)};
  \draw[black!60,line width=0.3pt]
    ({0.84*cos(\directionend/2)},{0.84*sin(\directionend/2)-0.025})
    -- ({0.84*cos(\directionend/2)},-0.25) -- (1.05,-0.25);
  \node[anchor=west,align=left] at (1.1,-0.15)
    {Human technical direction\\3 (0.3\%)};
\end{tikzpicture}
   \caption{Human messages by role ($N=1{,}071$).
  Areas show exact shares; percentages are rounded.}
  \label{fig:human-interaction-roles}
\end{wrapfigure}
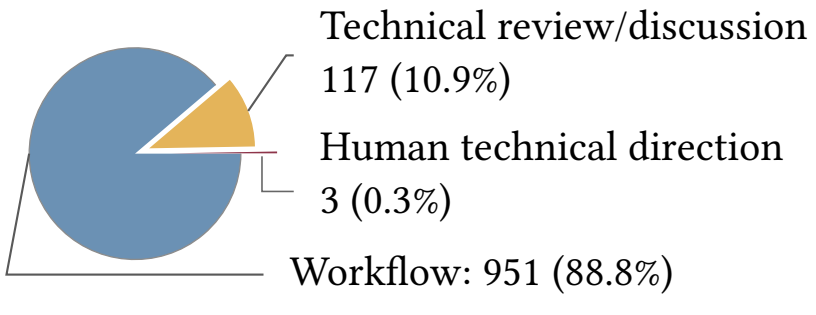

To characterize this involvement, we analyzed all 1,071 substantive human
messages exchanged across 211 sessions during the development leading to RC1.
Each message counts once, regardless of length; these interaction counts
are separate from the prompt-token total above.
Figure~\ref{fig:human-interaction-roles} summarizes the roles of these
human contributions.

Workflow organizes work through proof-task assignments, review scheduling,
phase transitions, or retry policies. Naming a theorem or asking agents
to prove it does not by itself supply technical content. Technical review
and discussion engage with that content, for example by asking why a
proof fails, examining assumptions, or approving an agent's proposal,
without introducing a substantive technical proposal. Human technical
direction supplies a concrete diagnosis, correction, implementation
alternative, or proof strategy. It need not override an agent's
recommendation or be correct. When roles overlap, technical direction
takes precedence over review and discussion, which takes precedence over
workflow.

Human contributions centered on coordinating parallel work (951 messages,
88.8\%) and reviewing or discussing technical content (117 messages,
10.9\%), including questions, investigation, and approvals. This active
coordination and review accompanied remarkably little human-supplied
technical direction: only three messages (0.3\%). The three directions comprise a
tentative proposition-repair suggestion subsequently corrected by the
agent, a conditional fallback to upstream register allocation, and a
later change from local prototype repairs toward a CompCert-style
allocator. The latter two belong to the same allocator-development
episode; Appendix~\ref{app:human-interventions} describes the corpus and
classification procedure and gives the context and follow-through for all
three cases.
 \FloatBarrier
\section{Evaluation}
\label{sec:evaluation}

\chen{We evaluate CompCert4Lean along three questions. Does the port
compile real programs correctly and as efficiently as CompCert
(\S\ref{sec:eval-release})? Is the port faithful to CompCert, pass by
pass (\S\ref{sec:eval-differential})? And do
coding agents improve the compiler further in Lean than in Rocq, while
keeping its correctness proof (Section~\ref{sec:eval-campaigns})?}

\subsection{\chen{Benchmarks and Release Results}}
\label{sec:eval-release}

\textbf{Benchmarks and correctness results.}
Our benchmark set is exhaustive within the scope defined in
Section~\ref{sec:overview-scope}. It includes (i)~\emph{all} CompCert~3.17
small-tests within that scope and (ii)~\emph{all} SPEC CPU2026 benchmarks
written entirely in C that fall within the C subset supported by CompCert.
Benchmark membership is determined by these scope and language criteria,
independently of measured compiler performance. Every qualifying program
is included, ruling out cherry-picking programs for favorable results
within these two suites.

For Table~\ref{tab:benchmark-coverage}, LoC counts physical lines in compiled
C translation units before preprocessing, including comments, blank lines,
and inactive code, but excluding headers and system libraries. The
small-tests mean covers its 29 performance programs and is rounded to the
nearest line; shared source files count once per program using them.
\chen{SPEC programs count once across inputs.}

\begin{table}[!htbp]
  \centering
  \caption{Benchmark coverage and recorded correctness results for
  CompCert4Lean \texttt{rc-0.0.1}. LoC reports physical source lines:
  the mean per program for small-tests and the total per program for SPEC.
  Correctness counts are per optimization profile; pass columns report
  passed/selected cases.}
  \label{tab:benchmark-coverage}
  \begingroup
  \small
  \setlength{\tabcolsep}{3pt}
  \begin{tabular*}{\linewidth}{@{\extracolsep{\fill}}llrrrrr@{}}
    \toprule
    Benchmark & Domain & LoC
      & \shortstack[r]{Correctness\\cases}
      & \shortstack[r]{Performance\\workloads}
      & \shortstack[r]{O0\\passed} & \shortstack[r]{O2\\passed} \\
    \midrule
    CompCert small-tests & Mixed & 2,909 (avg.) & 139 & 29 & 139/139 & 139/139 \\
    \midrule
    LBM & Fluid simulation & 1,016 & 2 & 2 & 2/2 & 2/2 \\
    Neutron & Particle transport & 3,832 & 4 & 4 & 4/4 & 4/4 \\
    TeaLeaf & Heat conduction & 4,284 & 1 & 1 & 1/1 & 1/1 \\
    NAB & Molecular modeling & 17,776 & 1 & 1 & 1/1 & 1/1 \\
    Zstd & Compression & 34,920 & 1 & 1 & 1/1 & 1/1 \\
    SQLite & Database & 232,586 & 1 & 1 & 1/1 & 1/1 \\
    \bottomrule
  \end{tabular*}
  \endgroup
  \vspace{-6pt}%
\end{table}

Table~\ref{tab:benchmark-coverage} summarizes the benchmark coverage and
correctness checks for the CompCert4Lean \texttt{rc-0.0.1} release on
Linux x86-64. The imported CompCert small-tests suite contributes 139
applicable correctness cases, covering language regressions, ABI
interoperability, numerical kernels, and applications. We treat all 29
performance workloads as one group. The nine documented
correctness exclusions concern other architectures, a frontend-only negative
test, the \chen{Rocq} export group, mixed-struct ABI modes excluded upstream on
x86-64, and a test whose expected result depends on non-portable pointer
arithmetic. The six
qualifying SPEC CPU2026 programs span numerical simulation, molecular
modeling, compression, and database processing. \chen{They compile
unmodified, using serial configurations and the required build flags.}

All included correctness cases pass at both O0 and O2; the pinned upstream
CompCert~3.17 also passes these checks. The six SPEC programs contribute ten
selected test/train inputs, giving 20 CompCert4Lean executions
across the two optimization profiles. Correctness uses the supplied
expected outputs or workload-specific checks, including SPEC's output
comparators. These execution and integration checks complement the
compiler's semantic-preservation proof. The 139-case small-tests count
includes compile-only checks, and its correctness and performance
populations overlap. The selected SPEC inputs are not official SPEC
performance scores.

\FloatBarrier
\textbf{Performance results.}
Table~\ref{tab:release-performance} reports generated-program runtime and
compiler execution cost with CompCert~3.17 fixed at a baseline of one.
Every entry divides the named compiler's time by the corresponding
CompCert time, so lower values are better. The program-runtime columns
compare O2 executables produced by CompCert4Lean \texttt{rc-0.0.1} and
GCC~15.2.0 against those produced by CompCert~3.17. The compilation
columns compare CompCert4Lean with CompCert at O0 and O2. Compilation
covers C-to-object translation, including preprocessing and assembly,
while excluding final linking and helper setup. We sum translation-unit
times within each program and take geometric means of the per-program
ratios for the 29 small-tests workloads. Runtime aggregates are geometric
means of per-input ratios, giving each input equal weight. On small-tests,
CompCert4Lean closely matches CompCert's generated-program runtime
(ratio 1.004), while GCC achieves 0.796; CompCert4Lean's compilation ratios
are 2.16 at O0 and 2.18 at O2.

The small-tests runtime measurements use the retained, hash-checked
release-validation executables and the retained GCC executables, with
one warmup and seven measured executions per compiler. Execution is
serial on logical CPU~2 of an AMD Ryzen~9 5900X, and compiler order rotates
between rounds. All 29 workloads contribute, including the shortest GCC
execution (approximately 9\,ms); timings include process startup.
The SPEC runtime values use the original quick-profile test/train inputs.
SPEC runtime and all compilation values are single-run observations. Small-tests
compilation used four concurrent case jobs, whereas SPEC compilation and
execution were serial. The LBM reference and GCC observations are from
September~23. The other SPEC reference observations are also from
September~23, Lean observations are from September~24, and the five
additional GCC observations are from October~3. These differences in
measurement protocol and date make
the compilation and SPEC values provisional; small runtime differences
should not be interpreted as statistically established improvements.

\begin{table}[htbp]
  \centering
  \caption{Recorded release performance, normalized to
  CompCert~3.17 $=1$ in every column. Lower is better: values below one
  indicate less time than CompCert. Program runtimes use O2; compilation ratios use the
  indicated optimization profile. The small-tests runtime row aggregates
  medians over all 29 workloads; SPEC runtime rows aggregate the selected
  test/train inputs. SPEC runtime and all compilation values are
  single-run observations, with the measurement qualifications described
  in the text.}
  \label{tab:release-performance}
  \begingroup
  \small
  \setlength{\tabcolsep}{3pt}
  \begin{tabular*}{\linewidth}{@{\extracolsep{\fill}}lrrrrr@{}}
    \toprule
    Benchmark & Inputs & \multicolumn{2}{c}{Program runtime (O2)}
      & \multicolumn{2}{c}{\shortstack{Compilation time\\(Lean/CompCert)}} \\
    \cmidrule(lr){3-4}\cmidrule(l){5-6}
      & & Lean/CompCert\,$\downarrow$ & GCC/CompCert\,$\downarrow$ & O0\,$\downarrow$ & O2\,$\downarrow$ \\
    \midrule
    CompCert small-tests & 29 & 1.004 & 0.796 & 2.16 & 2.18 \\
    \midrule
    LBM & 2 & 0.977 & 0.651 & 1.96 & 1.61 \\
    Neutron & 4 & 0.988 & 0.457 & 1.67 & 1.76 \\
    TeaLeaf & 1 & 0.966 & 0.629 & 1.52 & 1.59 \\
    NAB & 1 & 1.001 & 0.656 & 1.71 & 1.74 \\
    Zstd & 1 & 1.000 & 0.722 & 1.64 & 1.70 \\
    SQLite & 1 & 1.001 & 0.990 & 1.71 & 1.84 \\
    \bottomrule
  \end{tabular*}
  \endgroup
  \vspace{-6pt}%
\end{table}

\subsection{Differential Testing against CompCert}
\label{sec:eval-differential}

\chen{The correctness theorem relates CompCert4Lean to its own Lean
semantics, not to CompCert's Rocq development. We therefore compare
CompCert4Lean with CompCert~3.17 at three executable levels
(Table~\ref{tab:eval-levels}), using 22,062 test records from generated
programs, existing test suites, handwritten cases, bug replays, and
optimizer-divergence tests.}

\begin{table}[!htbp]
  \centering
  \caption{Differential-testing summary. Results concern the tested inputs;
  the output discrepancies comprise two undefined-behavior cases and
  35 mixed-compiler ABI cases.}
  \label{tab:eval-levels}
  \begingroup
  \small
  \setlength{\tabcolsep}{3pt}
  \begin{tabular*}{\linewidth}{@{\extracolsep{\fill}}lll@{}}
    \toprule
    Level & Comparison & Result \\
    \midrule
    Output & Exit status and stdout, O0/O2 & 37 discrepancies, explained \\
    IR & Complete RTL at nine stages & 18,807 (O2), 18,808 (O0) identical \\
    Specification & 202 executable functions & 37.4 million inputs, bit-identical \\
    \bottomrule
  \end{tabular*}
  \endgroup
  \vspace{-6pt}%
\end{table}

\textbf{\chen{Identical IR at every stage.}}
\chen{The strongest evidence comes from the intermediate representation.
On every one of the 18,807 programs compared at O2 (18,808 at O0), the
complete RTL is byte-identical after each of the nine RTL stages, with
not a single difference. Equal program output
alone would be weaker evidence, because a pass that was translated
incorrectly, for example a CSE that misses some redundancies, can still
produce correct programs. The IR comparison catches such a pass at its
own stage. To confirm that this check can fail, we mutated the Lean
passes: every pass mutant we built was detected, always at the stage of
the mutated pass (Appendix~\ref{app:differential-testing}).}

\chen{The other two levels agree. The 37 discrepancies
come from undefined behavior (2 records) and from structures passed to
GCC-compiled callees under CompCert's own calling convention (35
records); when caller and callee are compiled by the same compiler, these
programs agree across CompCert, CompCert4Lean, GCC, and Clang. At the
specification level, 202 executable functions give bit-identical results
on 37.4 million inputs, even for the floating-point operations that the
port defines with Lean's \texttt{Float.Model} instead of Flocq.}

\chen{These tests check the compiler, not the Lean semantics
or statements; their correspondence with the Rocq source is
checked by the source audits in Section~\ref{sec:approach-audit}. At the
IR level, tests also stop before register allocation, and the parser
and assembler are shared with CompCert
(Appendix~\ref{app:differential-limitations}).}
\FloatBarrier

\subsection{\chen{Optimization Campaigns}}
\label{sec:eval-campaigns}

The preceding experiments establish that CompCert4Lean produces executable
programs and closely follows the original compiler. We next ask whether
migration also makes this infrastructure more useful to coding agents.
Specifically, can an agent improve generated-program performance while
preserving the compiler's correctness guarantees, and does it achieve better
results in Lean than in Rocq? We evaluate this question through optimization
campaigns with GPT-6-astra, DeepSeek~4.1~Flash, and Kimi~K3.1,
using one campaign per model in each environment.

Each campaign starts from a frozen compiler version and asks the agent to
improve O2 program performance while retaining correctness and satisfying
compilation-time constraints. The agent explores optimization ideas, modifies
the compiler, checks its changes, and measures candidate performance. It may
revise existing optimization policies or introduce new transformations with
the proofs needed to integrate them into the compiler.
Table~\ref{tab:model-optimization-summary} summarizes the selected candidates
on 29 small-tests programs and ten SPEC inputs, with times normalized to
original CompCert~3.17. It also reports new verified passes
and retained source-line changes. The campaigns use a budget of 24 active hours. These are separate searches rather than repeated trials;
measurement qualifications therefore matter
when interpreting the comparison.
The Astra and DeepSeek measurements use different compression and SPASS
inputs across environments. We remeasure Kimi on identical inputs with a
\chen{shared, fresh original-CompCert reference.
Appendix~\ref{app:optimization} gives the protocols.}

New verified passes count distinct transformations
integrated into the selected compiler with a new correctness proof,
excluding heuristic changes and reuse of existing validators. Astra's Lean
count includes path splitting, loop unrolling, and instruction scheduling;
the first two share an implementation and proof family. Source lines changed
counts additions plus deletions in the final compiler implementation and proof
sources relative to each campaign's frozen baseline, including OCaml and
compiler diagnostics but excluding tests, scripts, documentation, and generated
files. \chen{Whitespace-only changes are ignored. This measures
retained change size, not difficulty.}

\begin{table}[htbp]
  \centering
  \caption{Summary of optimization campaigns. Each cell reports Rocq / Lean.
  O2 time ratios use original CompCert~3.17 as the baseline ($=1$; lower
  is better). Runtime geometric means cover 29 small-tests programs and
  ten SPEC inputs separately; compilation covers all 39 workloads.
  Measurement details appear in Appendix~\ref{app:optimization}.}
  \label{tab:model-optimization-summary}
  \begingroup
  \small
  \setlength{\tabcolsep}{2pt}
  \begin{tabular*}{\linewidth}{@{\extracolsep{\fill}}lrrrrr@{}}
    \toprule
    Model & \shortstack{Small-tests\\runtime} & \shortstack{SPEC\\runtime}
      & \shortstack{Compilation\\time}
      & \shortstack{New verified\\passes}
      & \shortstack{Source lines\\changed} \\
    \midrule
    GPT-6-astra & 0.977 / 0.883 & 0.978 / 0.931 & 1.054 / 2.054
      & 0 / 3 & 252 / 2,833 \\
    DeepSeek 4.1 Flash & 0.992 / 0.950 & 0.977 / 0.911 & 1.055 / 1.866
      & 0 / 0 & 128 / 136 \\
    Kimi K3.1 & 0.994 / 0.977 & 0.981 / 0.985 & 1.051 / 2.001
      & 0 / 0 & 27 / 23 \\
    \bottomrule
  \end{tabular*}
  \endgroup
  \vspace{-6pt}%
\end{table}

\textbf{Optimization results.}
All three models produce Lean candidates with lower small-tests runtime
than their Rocq counterparts; Astra and DeepSeek also achieve lower SPEC
runtime (Table~\ref{tab:model-optimization-summary}). For GPT-6-astra, the small-tests ratios are 0.883 in Lean and
0.977 in Rocq, while the SPEC ratios are 0.931 and 0.978. DeepSeek achieves
corresponding ratios of 0.950 versus 0.992 on small-tests and 0.911 versus
0.977 on SPEC. Kimi~K3.1 achieves small-tests ratios of 0.977 in Lean and
0.994 in Rocq. On SPEC, its ratios are 0.985 and 0.981, respectively;
the Lean candidate has a small advantage on small-tests, while the Rocq
candidate is slightly faster on SPEC.
The Lean
candidates retain higher compilation costs: Astra's compilation-time ratio
is 2.054, compared with 1.054 for Rocq, and DeepSeek's is 1.866, compared
with 1.055. Astra also produces 106 variants completing full compiler and
proof builds in Lean, compared with 40 in Rocq.

These results illustrate the practical benefit of bringing established
verification infrastructure into Lean, while also showing that the benefit
depends on how a model uses that infrastructure. Astra provides the clearest
example of extending the compiler: its selected Lean version introduces
path splitting, loop unrolling, and instruction scheduling, with the first
two sharing an implementation and correctness proof. It retains 2,833
changed source lines, compared with 252 in its Rocq version. DeepSeek adds
no new standalone verified pass, yet improves performance through new
inlining-selection and layout heuristics in Lean, retaining 136 changed
lines compared with 128 in Rocq. Kimi likewise retains heuristic changes,
with 23 changed source lines in Lean and 27 in Rocq, and no new verified
pass in either selected compiler. \chen{Thus, improvements arise from
new verified transformations and better use of existing passes. The
advantage supports the value of the migration for agent-driven
compiler development, although one campaign per model and environment
cannot separate model capability, proof infrastructure, repository
layout, and search variability.}

\textbf{GPT-6-astra's Lean optimization study.}
Beyond the aggregate results, Astra's new passes illustrate the scope of
autonomous development enabled by the migrated infrastructure. Given the
overall objective of improving generated-program performance, Astra
independently selected, designed, implemented, proved, and evaluated these
optimizations during the campaign of 24 active hours, without further human input.
It developed a general RTL code-copying transformation for both acyclic
path splitting and loop unrolling. Path splitting keeps execution paths
\chen{separate at joins so that constant propagation can specialize
them. Loop unrolling exposes successive iterations to existing optimizations
by copying the loop body and redirecting control flow between them.}

These transformations must preserve execution through branches, calls, and
returns. Astra constructed a simulation relating each source instruction to
its target copies, preserving register and memory contents while tracking
corresponding return addresses in the call stack. The proof supports
arbitrary routing policies, separating the correctness of the copying
machinery from the decisions about which copies execution visits. This
design allowed Astra to reuse the same implementation and proof for two
distinct optimization passes and integrate both into the compiler's
end-to-end correctness argument.

Astra also independently developed an RTL instruction scheduler that reorders
adjacent independent operations and loads before register allocation. A
liveness-based heuristic selects exchanges intended to reduce register
pressure, while dependency checks ensure that the instructions neither
consume each other's results nor write the same register. Control flow
introduces an additional challenge: another edge may enter the second
instruction directly. The implementation creates a separate block for the
reordered pair while preserving those existing entries, so that reordering
does not cause such executions to perform an extra instruction.

The scheduling proof accounts for intermediate states in which the source
and target have executed different instructions. Its simulation tracks this
partial progress until both reordered instructions have executed;
independence preserves their inputs, and updates to distinct registers
commute, restoring agreement. Astra integrated the resulting proof into the
verified compilation pipeline and evaluated the optimization as part of its
search. Together, these cases demonstrate autonomous development of new verified
extensions: the agent identified optimization opportunities and carried
them through algorithm design, implementation, proof construction, and
integration.

\textbf{Scope and interpretation.}
The retained changes implement reusable compiler optimizations, although
their selection and tuning are informed by the evaluation workloads.
Inspection of the Astra and DeepSeek candidates found no added conditions
that recognize benchmark names, application identities, source filenames,
or benchmark inputs. Instead, the optimizations apply through structural
and semantic properties, such as function size, static call count, loop
structure, instruction dependencies, and the presence of division operations.
Some policies are deliberately narrow: Astra's additional unrolling of
division-containing loops combines several size and control-flow
restrictions, while DeepSeek's inlining thresholds reflect performance and
compilation costs observed on the benchmarks. These changes can therefore
apply to other programs with the same properties, \chen{but their measured
benefits are not established improvements on unseen}
workloads.

Both environments permit workload-informed search under a common
optimization objective and constraints. The campaigns demonstrate
improvements to generated-program performance and integration of verified
extensions under these conditions. They characterize what agents achieved
in each environment, without establishing unseen-workload gains or
isolating the causes of environment differences.
Appendix~\ref{app:optimization} details the measurement qualifications.

\FloatBarrier
\section{Related Work}
\label{sec:related-work}

\textbf{Migrating formal developments between proof assistants.}
Prior work translates proofs between systems of the same logical family.
\citet{obua2006importing} import HOL4 theories into Isabelle/HOL by
replaying primitive inferences, \citet{kaliszyk2013scalable} make such
LCF-style translation scale to large libraries, and
OpenTheory~\cite{hurd2011opentheory} defines a shared exchange format
for higher-order logic theories. These approaches rely on a close
correspondence between the source and target logics. Within Lean, the
community ported Mathlib~\cite{mathlib2020} from Lean~3 to Lean~4 with
Mathport~\cite{mathport}, a semi-automatic translator followed by
extensive manual repair. Proof repair tools such as PUMPKIN
PATCH~\cite{ringer2018pumpkin} and repair across type
equivalences~\cite{ringer2021proofrepair} adapt proofs to changes in
definitions within one proof assistant. In contrast, Rocq 
and Lean differ
in their libraries, automation, and many representation choices, so a
syntactic translation does not carry proofs across. Our workflow
instead re-derives the development with LLM agents \chen{and uses
declaration accounting and independent audits to check that
translated definitions and statements still match the source.}

\chen{\textbf{Porting Rocq developments to Lean.}
The closest effort to ours is Iris-Lean~\cite{demedeiros2026irislean},
which ports the Iris separation-logic framework from Rocq to Lean~4 for
similar reasons, such as reusing Mathlib. Starting from a proof-mode
prototype~\cite{koenig2022iris}, the port took several years of work by
a large group of contributors; LLM agents entered only late, porting one
Iris-based development within a few days.
Other work automates parts of the
translation. Babel-formal~\cite{stoskopf2025babel} uses
proof terms as a pivot to help LLMs translate proof scripts between Lean
and Rocq, and \citet{viennot2025minif2f} use LLMs to translate the
miniF2F theorem statements into Rocq. 
ITPEval~\cite{wu2026itpeval}
benchmarks LLM translation across Lean, Rocq, Isabelle, and HOL Light,
and reports that its strongest evaluated model, GPT-5.5, achieves 5.2\%
proof-translation pass@1 (36 of 696 directed pairs) on the
Formalizing 100 Theorems tier drawn from real libraries. These
efforts either port a development by hand over years or translate
individual files, statements, and proofs. In contrast, CompCert4Lean
migrates a complete verified compiler, including its end-to-end
correctness theorem, with parallel LLM agents in about six weeks.}

\textbf{LLMs for theorem proving and verified code.}
Language models have been applied to formal theorem proving since
GPT-f~\cite{polu2020gptf}. Subsequent systems guide provers with informal
proof sketches~\cite{jiang2023dsp}, generate and repair whole
proofs~\cite{first2023baldur}, retrieve relevant premises in
Lean~\cite{yang2023leandojo}, and train on large synthetic proof
corpora~\cite{xin2024deepseekprover}. Related work generates code
together with specifications and proofs, for example in
Dafny~\cite{sun2024clover} and Verus~\cite{aggarwal2025alphaverus}.
\chen{These systems mostly target single theorems or small programs with
fixed statements. CompCert4Lean shows agent-driven migration at the scale
of a verified compiler: the workflow audits translated statements and
coordinates reconstruction of 4,908 interdependent obligations
across parallel agents sharing definitions and interfaces.}

\section{Conclusion}
\label{sec:conclusion}

CompCert4Lean demonstrates large-scale migration of a verified compiler
from Rocq to Lean by one expert directing parallel LLM agents over
approximately six weeks. Of 1,071 human interactions, 88.8\% concerned
workflow and coordination, while only three supplied technical direction.
The workflow combined dependency-based parallelism, correspondence audits,
and proof reconstruction against stable interfaces.

The optimization campaigns demonstrate a practical benefit of the Lean
port: under matched budgets, all three models achieved larger small-tests
runtime reductions in Lean while retaining the end-to-end correctness
theorem. Agents also integrated new verified extensions into the migrated
compiler. Together, these results suggest that agent-assisted migration
could make other Rocq-based tools available in Lean with manageable human
effort, while opening opportunities to build on and extend their verified
infrastructure.
 
\textbf{AI Use.}
AI tools powered the research agents described in
Sections~\ref{sec:approach} and~\ref{sec:implementation}; the optimization
experiments are documented in Section~\ref{sec:eval-campaigns}.
We also used AI tools to assist with editing the paper. The authors take
full responsibility for the paper's content and the research results
presented here.

\section*{Data-Availability Statement}
The CompCert4Lean source code is publicly available at
\url{https://github.com/fanlong/compcert4lean}.

\bibliographystyle{ACM-Reference-Format}
\bibliography{references}

\clearpage
\appendix
\section{Investigation Gates}
\label{app:investigation-gates}

Section~\ref{sec:approach-gates} describes the investigation process and
the representation gate (Gate~1). This appendix gives the floating-point
and memory/injection cases (Gates~2--3), followed by the directly
attributable effort for all three gates.

\textbf{Gate 2: floating-point values and rounding proofs.}
A direct substitution of Lean's logical floating-point type would lose
source-visible information. Its model canonicalizes NaNs, while CompCert's
binary32 and binary64 values retain NaN signs and payloads. These bits are
observable through conversions and memory. The selected x86 policy also
specifies which NaN operand supplies the result and how a signaling NaN is
quieted. Agreement on ordinary numerical values would therefore leave
part of the compiler's semantics unspecified.

The gate prototype placed a raw-word interface around the pure numerical
model. Schematically, its public carriers were:
\begin{quote}
\begin{verbatim}
structure RawFloat64 where
  bits : UInt64
structure RawFloat32 where
  bits : UInt32
\end{verbatim}
\end{quote}
For binary arithmetic, the wrapper first handled NaN operands using
CompCert's x86 selection and quieting rules. Otherwise it invoked the pure
numerical model and mapped an invalid-operation NaN to CompCert's specified
default. Negation and absolute value operated directly on the sign bit.
This gave an exact representation contract: the wrapper word must equal
the source value's \texttt{to\_bits} result, including signed zero and every
NaN payload. Native floating-point execution was unnecessary for this
theorem-bearing feasibility experiment.

The main challenge was the proof interface. CompCert's short
\texttt{mul2\_add} lemma, used in optimization proofs, relies on Flocq to
show that adding a value to itself agrees with multiplying it by two.
Simplifying the corresponding Lean operations reduced the finite case to
missing rounding lemmas. We built a reusable layer interpreting a
significand--exponent pair as the exact dyadic value $m\,2^e$ and proved
that full rounding preserves the identity
\[
  \operatorname{round}(2m,e)
  = \operatorname{round}(m,e+1).
\]
Unfolding addition and multiplication then reduced both operations to the
same rounded value. A checked negative example prevented an unsound
generalization: the internal \texttt{roundWithAccuracy} helper can return
different representations for $(2,100)$ and $(1,101)$, despite their equal
dyadic values, when its exponent precondition is violated. The proof
therefore stated that precondition for the helper and established the
unconditional law at the full normalization operation.

Subsequent gate work established independent nearest-even specifications,
conversions, square-root correctness, and connections to representative
memory and compiler consumers. The final independent audit checked a
134-module Lean dependency graph and obtained exact raw-output agreement
with pinned CompCert on 54,789 cases across 26 operation/width families.
The numerical proofs and the bit-level tests addressed complementary
obligations. The architectural decision was to separate the exact public
bit contract from the numerical implementation and its reusable proof
bridge; the later production translation remained subject to its own
correspondence audit.

\textbf{Gate 3: memory operations and injection preservation.}
CompCert's memory model must support both execution and relations between
the memories of different compiler stages. A memory injection maps a
source block to a target block and an offset. Matching the results of
allocation, loads, and stores does not establish that these operations
preserve the full injection relation.

The investigation kernel-checked a counterexample to a deliberately
weakened relation that required only agreement of mapped cells. Let two
distinct source blocks map to the same target cell:
\[
  \iota(b_1)=\iota(b_2)=(b_t,0).
\]
Initially, the two source cells and the target cell all contain zero, so
the weakened relation holds. Matching stores of one into $b_1$ and $b_t$
preserve their correspondence but invalidate the relation for the
untouched $b_2$. This small example explains the role of CompCert's
non-overlap requirement: it prevents a target update from interfering
with another live source location. The counterexample was an adversarial
representation probe, rather than a defect attributed to the source
compiler.

We chose a source-shaped functional memory view with explicit contents,
current and maximum permissions, and the next-block identifier, together
with its well-formedness invariants. A raw finite map was insufficient:
arbitrary entries need not respect the current/maximum permission
invariant or the next-block boundary. An optimized storage representation
would need to preserve these invariants and justify its observations
against the functional view. We retained all injection clauses, including
non-overlap, pointer representability, validity, and inverse permissions,
as well as the source rule allowing an undefined memory value to inject
into any target value.

Gate completion required general operation-preservation proofs and real
external-call cases, including allocation and copying, over a coherent
memory model. The initial zero-length-copy proof established only an
interface shape and was insufficient for acceptance. The completed gate
combined the stronger proofs with 9,029 differential observations across
20 partitions, including 8,192 deterministic operation sequences. This
established that the chosen representation supported the relational
reasoning needed by compiler simulations, in addition to reproducing
individual operations.

\textbf{Investigation effort and resulting decisions.}
Table~\ref{tab:gate-effort} reports directly attributable effort for these
three gates. We include the initial focused coinduction and floating-point
studies and the development, integration, retry, and audit workers
assigned to each gate. Time is accumulated agent runtime, including tools
and waiting within a run; parallel workers contribute separately. The
shared 30-risk discovery campaign and shared coordination are excluded,
so these figures measure gate-specific work building on existing evidence.
In particular, Gate~1's short run reused the preliminary representation
studies, and later behavior/world consumer work remained charged to its
recorded Gate~3 assignments. The full phase cost appears in
Table~\ref{tab:phase-token-usage}.

\begin{table}[htbp]
  \centering
  \caption{Recorded effort directly attributed to Gates~1--3. Time is
  accumulated agent-hours, not elapsed calendar time. Token counts are in
  millions, rounded independently; cached input is included in Total,
  and output includes reasoning. Worker sessions are deduplicated and
  recorded unsuccessful attempts are included. Six Gate~2 worker sessions
  have no positive token counters, so usage is not a complete billing
  measurement. Shared discovery and coordination are excluded.}
  \label{tab:gate-effort}
  \begin{tabular*}{\linewidth}{@{\extracolsep{\fill}}lrrrrr@{}}
    \toprule
    Gate & \shortstack[r]{Agent\\hours}
      & \shortstack[r]{Uncached\\input (M)}
      & \shortstack[r]{Cached\\input (M)}
      & Output (M) & Total (M) \\
    \midrule
    1: Representations & 1.4 & 0.9 & 37.7 & 0.2 & 38.7 \\
    2: Floating point & 107.9 & 58.4 & 3,473.3 & 10.3 & 3,542.1 \\
    3: Memory/injections & 20.7 & 10.8 & 508.0 & 2.7 & 521.5 \\
    \bottomrule
  \end{tabular*}
\end{table}

 \clearpage
\section{Review Across Audit Rounds}
\label{app:audit-rounds}

Section~\ref{sec:approach-audit} describes the iterative audit and repair
procedure. This appendix reports the findings across review rounds and
the evidence and limitations of the final acceptance decision.

The first global audit exposed defects that translation-time integration
had missed. The build root imported none of the translated modules, so
the apparent whole-project build compiled only the root. The audit also
found 1,091 theorem or generated-fact propositions reduced to
\texttt{True}. We restored build aggregation and reconstructed the
propositions before relying on subsequent checks. Other findings
included missing transition premises, hidden configuration defaults,
and mismatched providers. Source review was complemented by differential
probes, downstream consumer examples, successful executions, and
negative validator tests. These checks supply different kinds of
correspondence evidence; finite tests do not establish universal
equivalence between the developments.

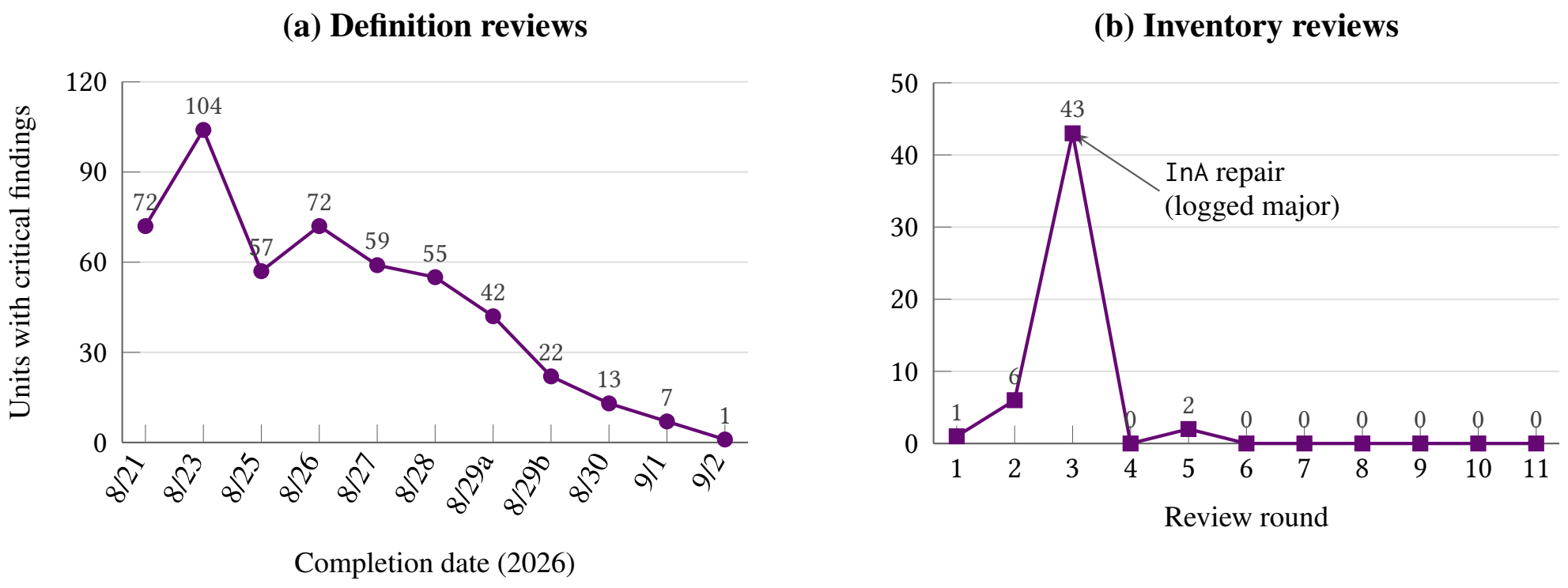
\begin{figure}[htbp]
  \centering
  \begin{minipage}[t]{0.49\linewidth}
    \centering
    \begin{tikzpicture}[baseline=(current bounding box.north)]
      \begin{axis}[
        width=\linewidth, height=4.6cm,
        title={(a) Definition reviews},
        title style={font=\figfont\bfseries\footnotesize},
        xlabel={Completion date (2026)},
        ylabel={Units with critical findings},
        label style={font=\figfont\scriptsize},
        ticklabel style={font=\figfont\scriptsize},
        xmin=0.6, xmax=11.4, ymin=0, ymax=120,
        xtick={1,...,11},
        xticklabels={8/21,8/23,8/25,8/26,8/27,8/28,8/29a,8/29b,8/30,9/1,9/2},
        x tick label style={rotate=60, anchor=east},
        ytick={0,30,60,90,120},
        axis x line*=bottom, axis y line*=left,
        axis line style={black!60}, tick style={black!60},
        ymajorgrids, grid style={black!12},
        clip=false,
        nodes near coords,
        every node near coord/.append style={
          font=\figfont\tiny, text=black!75},
      ]
        \addplot[thupurple, thick, mark=*, mark size=1.5pt]
          coordinates {
            (1,72) (2,104) (3,57) (4,72) (5,59) (6,55)
            (7,42) (8,22) (9,13) (10,7) (11,1)
          };
      \end{axis}
    \end{tikzpicture}
  \end{minipage}\hfill
  \begin{minipage}[t]{0.49\linewidth}
    \centering
    \begin{tikzpicture}[baseline=(current bounding box.north)]
      \begin{axis}[
        width=\linewidth, height=4.6cm,
        title={(b) Inventory reviews},
        title style={font=\figfont\bfseries\footnotesize},
        xlabel={Review round},
        label style={font=\figfont\scriptsize},
        ticklabel style={font=\figfont\scriptsize},
        xmin=0.6, xmax=11.4, ymin=0, ymax=50,
        xtick={1,...,11},
        ytick={0,10,20,30,40,50},
        axis x line*=bottom, axis y line*=left,
        axis line style={black!60}, tick style={black!60},
        ymajorgrids, grid style={black!12},
        clip=false,
        nodes near coords,
        every node near coord/.append style={
          font=\figfont\tiny, text=black!75},
      ]
        \addplot[thupurple, thick, mark=square*, mark size=1.5pt]
          coordinates {
            (1,1) (2,6) (3,43) (4,0) (5,2) (6,0)
            (7,0) (8,0) (9,0) (10,0) (11,0)
          };
        \node[font=\figfont\scriptsize, align=left, anchor=west,
          fill=white, inner sep=1pt] (ina) at (axis cs:4.5,35)
          {\texttt{InA} repair\\(logged major)};
        \draw[-{Stealth[length=1.3mm]}, black!65, thin]
          (ina.west) -- (axis cs:3,43);
      \end{axis}
    \end{tikzpicture}
  \end{minipage}
  \caption{Units with at least one finding classified as critical in
  successive definition and proof-inventory reviews. The phases use
  different severity rules; panel~(b) numbers completed inventory reviews
  consecutively from 1 and uses globally reconciled findings, with its
  round~3 peak dominated by missing authenticated evidence. The
  \texttt{InA} annotation marks the round of the case discussed in
  Section~\ref{sec:approach-audit}; that finding was classified as major. The first definition
  review covered 167 reporting units and subsequent reviews covered 166;
  8/29a and 8/29b distinguish two reviews completed on August~29.
  Inventory rounds cover 166 units using fresh reviews and, from round~5,
  unchanged evidence carried forward. Incomplete rounds and targeted
  rechecks are omitted. Counts can include findings repeated across rounds.}
  \Description{Two line charts show units with critical findings. Across
  eleven broad definition reviews from August 21 to September 2, the
  counts are 72, 104, 57, 72, 59, 55, 42, 22, 13, 7, and 1. Across
  inventory rounds 1 through 11, the counts are 1, 6, 43, 0, 2, and six
  zeros. An annotation at round 3 identifies the
  InA repair, separately classified as major.}
  \label{fig:audit-critical-units}
\end{figure}

Figure~\ref{fig:audit-critical-units} shows the number of units with
critical findings across successive reviews. Across eleven completed
broad definition reviews, the count peaked at 104 and fell to one in the
last broad round. A targeted recheck of two repaired units then closed
the remaining findings. Subsequent inventory reviews continued to expose
missing obligations and interface mismatches. The counts include missing
correspondence evidence as well as confirmed defects, and the same issue
can appear in more than one round.

Acceptance at line~20 of Algorithm~\ref{alg:definition-migration} is the decision to use the reviewed definitions
and statements as fixed targets for ordinary proof work. The inventory
records their intended Lean types, source correspondence, responsible
declarations, and known proof dependencies. Our accepted inventory
contained 4,908 targets, including theorem proofs and proof-bearing
certificates; 170 were already complete. The remaining duties become
inputs to Algorithm~\ref{alg:proof-grinding}. A snapshot taken for one
audit round can still be repaired in the next round; the definition
freeze for proof work begins only with this acceptance decision.

The returned evidence and decision also delimit what acceptance
establishes. In our final obligation-inventory round, 25 units lacked an
accepted result, and the expert waived the planned complete-cycle and
external final-global gates after a bounded repair and targeted recheck.
Those gates were not reported as passed. Acceptance records this
engineering judgment and its residual risks; it neither discharges the
remaining proofs nor establishes mechanized Rocq--Lean equivalence.

 \clearpage
\section{Automated Proof Reconstruction and Proof Progress}
\label{app:proof-progress}

Section~\ref{sec:approach-proofs} describes automatic proof reconstruction.
The following milestones record its progress and the subsequent completion
of the remaining obligations.

\textbf{Automated proof reconstruction and proof progress.}
Table~\ref{tab:proof-progress} records these milestones in our
migration. At the expert-directed stop, the runner recorded 4,267 new
completions in addition to the 170 baseline witnesses: 90.1\% of the
initially unfinished 4,738 targets had been completed. The remaining 471
included 161 targets in integer foundations, 36 in value-domain reasoning,
and 23 in memory, as well as compiler-pass proofs. The residual workload
therefore called for investigations across shared foundations and their
consumers, rather than only the final assembly of large simulations.

\begin{table}[htbp]
  \centering
  \caption{Checked proof-inventory milestones. Counts include the 170
  initially complete targets. Automatic and guided work overlapped;
  these are cumulative milestones, not results from disjoint phases or
  equal effort budgets. The automatic runner was stopped by the expert
  with unfinished work still queued.}
  \label{tab:proof-progress}
  \begin{tabular*}{\linewidth}{@{\extracolsep{\fill}}lrr@{}}
    \toprule
    Milestone & Completed & Unfinished \\
    \midrule
    Accepted definition baseline & 170 & 4,738 \\
    Automatic runner stopped & 4,437 & 471 \\
    Later transition to simple status ledger & 4,732 & 176 \\
    Final integrated proof baseline & 4,908 & 0 \\
    \bottomrule
  \end{tabular*}
\end{table}

We simplify coordination when the remaining workload no longer
benefits from the automatic campaign's machinery. After further guided
work had raised completion to 4,732 targets, we replaced the campaign
ledger with a manually maintained boolean status ledger for the remaining
176 obligations (Table~\ref{tab:proof-progress}).
The earlier inventory and execution history were preserved as references,
while integration retained the requirements of \textsc{ValidExtension}
in Algorithm~\ref{alg:proof-grinding}. The simpler ledger changed how
work was assigned and tracked; finished status still required checked
integration, and definition repairs still required explicit review.
This process completed all 4,908 recorded targets, including the
supporting certificates, pass simulations, and composed
compiler-correctness result.
 \clearpage
\section{Localized Floating-Point Repairs}
\label{app:float-repairs}

Section~\ref{sec:approach-finalize} presents the missing-format-assumptions
repair required during human-guided proof finalization. This appendix
gives the counterexample, repair, and subsequent proof checks for the
other case, a conversion that omitted destination normalization.

\textbf{A conversion that lost the value one.}
Four conversion obligations were false under the migrated implementation.
A checked witness starts with single-precision $1.0$, whose bit pattern
is 1,065,353,216. Widening it produced the double-precision bit pattern
8,388,608, and converting that result to an integer returned zero instead
of one. The reverse route also failed: converting integer one directly
to single precision produced $1.0$, whereas converting through double
precision and narrowing produced zero. Agents proved the negations of
the four affected propositions in Lean, separating a definition defect
from a difficult proof.

Both conversion functions unpacked a source-format finite value and
passed it directly to the destination format's packing operation.
Although the mantissa and exponent represented the source value, packing
required them to be normalized for the destination format. The authorized
repair inserted destination normalization before packing, including
round-to-nearest-even for narrowing, while retaining the interfaces and
the treatment of NaNs, zeros, and infinities. Rechecking the conversion
proofs then established the intended propositions. This case shows why
successful proofs of surrounding operations cannot replace checking the
representation invariants at a format boundary.
 \clearpage
\section{Differential Testing Details}
\label{app:differential-testing}

This appendix gives the corpus, comparison procedures, detailed outcomes,
and limitations underlying Section~\ref{sec:eval-differential}.
The corpus counts records and distinct source sets, whereas the IR outcome
table counts distinct source-and-flags programs; per-file stage pairs and
specification inputs are separate units of comparison.

\textbf{Test corpus.}
The first two levels share one corpus of 22,062 test
records (18,999 distinct source sets) from 18 campaigns
(Table~\ref{tab:eval-corpus}): random programs (Csmith~\cite{yang2011csmith}
with its default, feature, floating-point, and large profiles,
YARPGen~\cite{livinskii2020yarpgen}, and EMI
mutants~\cite{le2014emi}), existing test suites (the GCC torture tests,
the LLVM test suite, and the CompCert test suite), handwritten tests of
integer, floating-point, ABI, and control-flow semantics, replays of
known CompCert bugs, and programs selected to make the two optimizers
diverge. We screen for undefined behavior by also compiling every record with GCC
and Clang with UBSan and ASan. A record with a sanitizer report is
classified as undefined and is not counted as correspondence evidence;
the absence of a report does not establish that a program is free of
undefined behavior.

\begin{table}[htbp]
  \centering
  \caption{Corpus by source and selected final output-level outcomes.
  ``Agree'': the four CompCert and CompCert4Lean executables (O0, O2) give
  the same exit status and output, with sanitizers clean or their run
  incomplete; this excludes the 362 records that agree with each other
  but differ from a reference. ``UB'': a sanitizer reports undefined
  behavior. ``Differ'': the outputs of the two compilers differ on a
  program without a sanitizer report. The displayed outcomes are partial:
  the remaining 2,923 records are classified in the full outcome table
  (Table~\ref{tab:eval-output}).}
  \label{tab:eval-corpus}
  \begingroup
  \small
  \setlength{\tabcolsep}{3pt}
  \begin{tabular*}{\linewidth}{@{\extracolsep{\fill}}lrrrr@{}}
    \toprule
    Source & Records & Agree & UB & Differ \\
    \midrule
    Random (Csmith, YARPGen, EMI)                 & 10,289 &  8,997 & 403 &  0 \\
    Test suites (GCC, LLVM, CompCert)             &  4,153 &  2,579 &  83 &  0 \\
    Handwritten (semantics, ABI, control, allocator) & 2,069 & 1,897 & 16 & 20 \\
    Bug replays and divergence mining            &  5,551 &  5,051 &  78 & 15 \\
    \midrule
    Total                                         & 22,062 & 18,524 & 580 & 35 \\
    \bottomrule
  \end{tabular*}
  \endgroup
\end{table}

\textbf{Output-level comparison.}
Our first comparison is end to end. One deterministic harness compiles
each record with both compilers at O0 and O2, with the same flags,
compile deadline (600\,s), link mode (non-PIE), and runtime library,
runs the four executables on the same input, and compares their exit
status and complete standard output. This test covers the complete
pipeline, including register allocation and the passes after it.
However, equal output is weaker evidence than it appears. Because both
compilers run the same passes in the same order, a pass that was
translated incorrectly but still preserves semantics, for example a CSE
that misses some redundancies, does not change the output.

\textbf{IR-level comparison.}
To expose such defects, we also compare the RTL program that each
compiler produces after each of its nine RTL stages (RTLgen, Tailcall,
Inlining, Renumber, Constprop, Renumber, CSE, Deadcode, Unusedglob). We
stop before register allocation, because CompCert4Lean replaces
CompCert's unverified OCaml allocator with its own allocator, which
legitimately chooses different registers. The Lean compiler encodes its
RTL as JSON, and the bridge decodes it into CompCert's own OCaml data
type. Both sides are then printed by the same printer, an extended
version of CompCert's \texttt{PrintRTL} that prints every field of the
program: all global
definitions in order, function signatures, parameters, stack sizes, entry
points, every instruction with its full call and builtin signatures,
external functions, global variables with their initialization data and
flags, \texttt{prog\_public}, and \texttt{prog\_main}. Every
floating-point constant is printed as its exact bit pattern. The printer
has no wildcard patterns, and OCaml's exhaustiveness check confirms that
it omits no constructor or field. Equality after stage~0 checks the Lean
passes from SimplExpr to RTLgen as one block, and equality after each
later stage checks one RTL pass. At O0 only stages 0, 3, and~8 change the
program, so O0 checks the front end, Renumber, and Unusedglob; at O2 every
pass changes most programs (Table~\ref{tab:eval-passes}).

One gap remains: the dumps come from the diagnostic driver of
CompCert4Lean, not from the proven driver. To narrow this gap, we proved
in Lean that the diagnostic driver with a
silent sink is equal to the proven driver \texttt{transf\_c\_program}
(theorem \texttt{transf\_c\_program\_silent}, which depends only on
\texttt{propext}, \texttt{Classical.choice}, and \texttt{Quot.sound}).

\textbf{Sensitivity of the IR check.}
A check that never reports a difference is convincing only if it can
report one. To confirm that it can, we built mutants of the Lean passes
and of the RTL fields. Each pass mutant changes only a definition,
so no proof had to change. Every tested pass mutant was detected on at least one program, and each
detection was at the stage of the mutated pass (Table~\ref{tab:eval-passes}). The
detection rate per program is lower for mutants that only weaken an
optimization (27, 39, and 43 of 180 programs for Constprop, CSE, and
Deadcode), and most detections come from Csmith programs. No Tailcall mutant was tested. The field
mutants, which change data that the original \texttt{PrintRTL} does not
print, are detected on every program that they change.

Pass mutants in Table~\ref{tab:eval-passes} were evaluated with the
original \texttt{PrintRTL}.

\begin{table}[htbp]
  \centering
  \caption{RTL pass activity and mutation detection at O2. ``Changed'' reports
  the percentage of 20,572 records producing RTL in both compilers on
  which the pass changes the program; duplicates are included.
  ``Non-gen.'' uses the 4,812 non-randomly-generated records in that set.
  ``Detected'' reports detections out of 180 tested programs.
  The single Renumber miss is a timeout.}
  \label{tab:eval-passes}
  \begingroup
  \small
  \setlength{\tabcolsep}{3pt}
  \begin{tabular*}{\linewidth}{@{\extracolsep{\fill}}rllrrr@{}}
    \toprule
    & & & \multicolumn{2}{c}{Changed (\%)} & \\
    \cmidrule(lr){4-5}
    Stage & Pass & Mutation & All & Non-gen. & Detected \\
    \midrule
    0 & RTLgen     & branch heuristic \texttt{>} to \texttt{>=} & -- & -- & 55/180 \\
    1 & Tailcall   & --                                         & 63.9 & 23.9 & -- \\
    2 & Inlining   & wrong called-once test                     & 67.1 & 28.4 & 70/180 \\
    3 & Renumber   & reverse postorder numbering                & 94.0 & 85.0 & 179/180 \\
    4 & Constprop  & no condition strength reduction            & 93.0 & 75.0 & 27/180 \\
    5 & Renumber   & (as stage 3)                               & 87.6 & 62.9 & -- \\
    6 & CSE        & a store kills all loads                    & 84.2 & 50.9 & 39/180 \\
    7 & Deadcode   & never remove dead stores                   & 89.0 & 64.6 & 43/180 \\
    8 & Unusedglob & keep all definitions                       & 66.9 & 28.0 & 72/180 \\
    \midrule
    \multicolumn{3}{@{}l}{Field mutants (extended printer, 16 programs, O0 and O2)} & & & \\
    \multicolumn{3}{@{}l}{\quad stack size, signatures, \texttt{prog\_public}, \texttt{prog\_main}} & & & 16/16 \\
    \multicolumn{3}{@{}l}{\quad global init data, volatile flag, definition order} & & & 14/16 \\
    \multicolumn{3}{@{}l}{\quad float constant $+1$ ulp (180 programs, O2)} & & & 14/180 \\
    \bottomrule
  \end{tabular*}
  \endgroup
\end{table}

\textbf{Specification-level comparison.}
The first two levels exercise the specification only through the
programs in the corpus. We therefore also test its executable functions
directly. We compared 202 executable functions of the Lean
specification with the OCaml code that Rocq extracts from CompCert~3.17:
the integer
operations of \texttt{Int}, \texttt{Int64}, \texttt{Byte}, and
\texttt{Ptrofs}; the operations and conversions of \texttt{Float} and
\texttt{Float32}; \texttt{sem\_cast} on all 576 pairs of types and the
unary, binary, and comparison operations of \texttt{Cop} on non-pointer
values; \texttt{encode\_val}, \texttt{decode\_val}, and
\texttt{proj\_bytes} of \texttt{Memdata} for all 11 memory chunks; and the
15 functions of \texttt{Values.Val} that the extraction keeps. The inputs
were generated with a fixed seed and biased to edge values: NaN payloads,
signed zeros, infinities, subnormals, rounding ties, and conversion
bounds. Two deliberate perturbations, one on each side, were detected.

\textbf{Reproducibility.}
All runs use one script per level and fixed seeds. A full rerun of
the output-level test assigns the same outcome to 22,029 of the 22,062
records (99.85\%). The 33 changes come from timing near a deadline and
from transient \texttt{ETXTBSY} errors in the harness; no program output
changed. We did not rerun the 18,875 records that agreed in that run
with the final harness. Instead, we reran a stratified sample of 2,000
of them with the final compiler and harness: 1,998 agree, and the
other~2 agree after a rerun at lower machine load, since their run time
is close to their 30\,s deadline.

\textbf{Results: behavior.}
\chen{Table~\ref{tab:eval-output} gives the final
outcome of every record.} The 35 differences in Table~\ref{tab:eval-corpus}
exclude the two undefined-behavior cases included in the 37 discrepancies
discussed below. Although the two compilers give different
results on 37 records, none remains an unexplained port discrepancy. Two have
undefined behavior. One is the
replay of CompCert issue~\#579, which uses the reserved identifier
\texttt{\_\_stringlit\_1}: CompCert~3.17 silently produces wrong output,
whereas CompCert4Lean rejects the program. The other is a GCC test on
which ASan reports a stack use after return. The remaining 35 records (20
distinct programs), which have no sanitizer report, pass structures by value
to an object file compiled by GCC. Both compilers pass structures with
CompCert's own x86-64 convention, which differs from the System~V
convention that GCC follows, so the GCC-compiled callee reads its argument from locations
that the caller did not write, and the output depends on register and
stack contents, which differ because the register allocators differ. When
the callee is compiled by the same compiler as the caller, all 20
programs agree across CompCert, CompCert4Lean, GCC, and Clang. There is
no record that CompCert rejects and CompCert4Lean accepts, and every
timeout and every compile failure other than \#579 is the same for both
compilers.

\begin{table}[htbp]
  \centering
  \caption{Final output-level outcome of the 22,062 records. Every
  compile failure and every run timeout of CompCert or CompCert4Lean,
  except \#579, is the same for both compilers at the same level.}
  \label{tab:eval-output}
  \begingroup
  \small
  \setlength{\tabcolsep}{3pt}
  \begin{tabular*}{\linewidth}{@{\extracolsep{\fill}}lr@{}}
    \toprule
    Outcome & Records \\
    \midrule
    Agree, sanitizers clean                                        & 18,489 \\
    Agree, sanitizer run incomplete                                &     35 \\
    Agree with each other, differ from a reference$^{a}$           &    362 \\
    Undefined behavior reported by a sanitizer$^{b}$               &    580 \\
    Run timeout (1,096 both compilers, 1 GCC only)                 &  1,097 \\
    Rejected by both compilers                                     &  1,423 \\
    Same failure pattern on both sides (e.g.\ both reject O0 only) &     40 \\
    \midrule
    Rejected by CompCert4Lean only (\#579, undefined behavior)    &      1 \\
    Different output, struct passing to GCC code (20 programs)    &     35 \\
    Rejected by CompCert only                                      &      0 \\
    \midrule
    Total                                                          & 22,062 \\
    \bottomrule
  \end{tabular*}
  \endgroup
  \\[2pt]
  {\footnotesize $^{a}$GCC or Clang (306), the CompCert interpreter (30), or
  the expected output (26). $^{b}$On one of them the outputs of the two
  compilers also differ (ASan: stack use after return).}
\end{table}

\textbf{Results: identical IR at every stage.}
The IR-level test gives a stronger result for the passes up to RTL
(Table~\ref{tab:eval-ir-diff}). Of the 20,279 distinct programs (by
source and flags), 18,808 at O0 and 18,807 at O2 produce RTL on both
sides. The complete RTL is identical at all nine stages for every comparable
program, including at O2.
The comparison runs per C file, and 1,059 of the compared programs have
more than one C file (most have two; the largest has 51). The compared
programs thus contain 19,938 C files at O0 and 19,937 at O2, which give
179,442 and 179,433 identical stage pairs, the largest dump several
megabytes. Another 59 C files, from the 59 programs that are only
partially comparable (Table~\ref{tab:eval-ir-diff}), also have identical
RTL at all nine stages (531 stage pairs per level). In total,
359,937 per-file stage pairs are identical and none differs. This total
overstates the number of independent pass checks: at O0 six of the nine
stages do not change the program (see the IR-level comparison above), so we report the result by program in
Section~\ref{sec:eval-differential}.
The remaining outcomes provide no complete pair to compare.
This is the result that a faithful
translation must give: the passes were translated function by function
from the Rocq source, with ports of CompCert's own maps, sets, and
integer arithmetic, so any difference would indicate a translation
defect.

\begin{table}[htbp]
  \centering
  \caption{IR-level differential testing of CompCert4Lean against
  CompCert~3.17 on 20,279 distinct programs. A program is compared when
  both compilers produce RTL; its complete RTL after each of the nine
  stages is then compared byte for byte.}
  \label{tab:eval-ir-diff}
  \begingroup
  \small
  \setlength{\tabcolsep}{3pt}
  \begin{tabular*}{\linewidth}{@{\extracolsep{\fill}}lrr@{}}
    \toprule
    Outcome & O0 & O2 \\
    \midrule
    Compared, identical at all nine stages & 18,808 & 18,807 \\
    Compared, different at some stage      &      0 &      0 \\
    \midrule
    Rejected by both compilers             &  1,411 &  1,411 \\
    Contains a precompiled object (C units identical) & 58 & 58 \\
    One C file rejected by both, other file identical &  1 &  1 \\
    Rejected by CompCert4Lean only (\#579, undefined behavior) & 1 & 1 \\
    Compile deadline (300\,s) exceeded by both &  0 &  1 \\
    \midrule
    Total                                  & 20,279 & 20,279 \\
    \bottomrule
  \end{tabular*}
  \endgroup
\end{table}

\textbf{Results: specification.}
On 37,400,000 inputs, all 202 functions give bit-identical results
on both sides. For one group of functions, this agreement was not
guaranteed by construction. The port defines floating-point addition,
subtraction, multiplication, division, and square root with the
IEEE~754 model of the Lean core library (\texttt{Float.Model}), not with
the Flocq functions that CompCert uses, and we found no theorem in the
port that relates the two. Even so, the two implementations agree on
every tested input, including the edge values above. For these
operations, the test is the only evidence that the Lean specification
agrees with CompCert's. \texttt{Float32.sqrt} is not in the
extracted code and was not tested.

\textbf{Limitations.}
\label{app:differential-limitations}
Despite these results, the tests have five limits. First, they do not execute the Lean
semantics (\texttt{Csem}, \texttt{Asm}) as a whole, and they give no
evidence about the proofs; a \texttt{Csem} rule that makes too many
programs undefined would not change any test result. Second, the C parser
and elaborator, the assembly printer, and the assembler are shared code,
so no test can expose a defect that the two compilers share there.
Third, register allocation and the passes after it (Allocation,
Tunneling, Linearize, CleanupLabels, Stacking, Asmgen) are checked only
by the output-level test, on inputs that differ from CompCert's because
the allocators differ; Debugvar is not run. Fourth, the IR comparison of
the Lean side passes through the JSON codec and a shared printer, and the
silent-sink theorem does not cover the recording sink that writes the
dumps; that the $k$-th dump is the $k$-th intermediate program follows
from the code of the driver. Fifth, the corpus rarely exercises
\texttt{switch} with jump tables, variadic functions, function pointers,
\texttt{setjmp}, and inlining of functions with more than one call
site.
 \clearpage
\section{Detailed Optimization Campaign Results}
\label{app:optimization}

Table~\ref{tab:optimization-progress} records the retained optimization
checkpoints for both compilers. Each entry is the geometric mean of 39
individual O2 time ratios, comprising 29 small-tests programs and ten SPEC
inputs, relative to original CompCert~3.17. For compilation, C-to-object times
of a program's translation units are summed before taking its ratio; the
program's ratio is repeated for each of its SPEC inputs. The CompCert4Lean
entries combine the archived release-validation CompCert reference times with
the measured times at each Lean checkpoint. The CompCert entries use the
original 3.17 measurements paired with each candidate in the \chen{Rocq} campaign.
These are measurements from separate campaigns, not a simultaneous paired
cross-compiler run. Moreover, the Lean campaign used a repeated 1\,MiB input
for the three compression programs and \texttt{problem.dfg} for SPASS, whereas
the \chen{Rocq} campaign used the original license input and
\texttt{small\_problem.dfg}. The compilation figures exclude linking and
helper setup. Values below one indicate less time than original CompCert.

\begin{table}[htbp]
  \centering
  \caption{GPT-6-astra optimization checkpoints relative to original
  CompCert~3.17. Each numeric entry is the geometric mean of 39
  per-workload O2 time ratios ($=1$; lower is better). Within each compiler
  group and column, the best result is in \textbf{bold} and the second best
  is \underline{underlined}; reference rows are not ranked.}
  \label{tab:optimization-progress}
  \begingroup
  \small
  \setlength{\tabcolsep}{3pt}
  \begin{tabular*}{\linewidth}{@{\extracolsep{\fill}}llrr@{}}
    \toprule
    Compiler version & Change & \shortstack[r]{O2 program\\runtime}
      & \shortstack[r]{O2 compilation\\time} \\
    \midrule
    Original CompCert~3.17 & Baseline & 1.000 & 1.000 \\
    CompCert~3.17 \texttt{d324ee55} & Block layout, spill costs & 0.993 & \textbf{1.011} \\
    CompCert~3.17 \texttt{fed9846f} & Small-leaf inlining & \underline{0.989} & \textbf{1.011} \\
    CompCert~3.17 \texttt{2b22ad49} & Small-wrapper inlining & \textbf{0.977} & \underline{1.054} \\
    \midrule
    CompCert4Lean \texttt{801facb} & Release version & 1.000 & 2.046 \\
    CompCert4Lean \texttt{3c06641} & Campaign baseline & 0.982 & \textbf{1.728} \\
    CompCert4Lean \texttt{e691e83} & Small-leaf inlining & 0.971 & \underline{1.800} \\
    CompCert4Lean \texttt{edacc68} & Four-copy loops & 0.935 & 1.955 \\
    CompCert4Lean \texttt{3cad2da} & Wider inlining, if-conversion & 0.922 & 1.961 \\
    CompCert4Lean \texttt{7057444} & Acyclic path splitting & 0.921 & 1.991 \\
    CompCert4Lean \texttt{d70b047} & Spill reload reuse & 0.916 & 1.961 \\
    CompCert4Lean \texttt{28fba67} & Division loops, reload cache & 0.911 & 2.011 \\
    CompCert4Lean \texttt{8dc9f4e} & Non-leaf inlining & 0.902 & 1.945 \\
    CompCert4Lean \texttt{558a647} & RTL instruction scheduling & \underline{0.897} & 2.056 \\
    CompCert4Lean \texttt{742233e} & Loop-weighted spill costs & \textbf{0.895} & 2.054 \\
    \bottomrule
  \end{tabular*}
  \endgroup
\end{table}

Table~\ref{tab:optimized-case-results} reports the best retained versions of
both compilers at O2: CompCert~3.17 \texttt{2b22ad49} and CompCert4Lean
\texttt{742233e}. Each value is a time ratio to original CompCert~3.17, with
smaller values indicating less time. The small-tests row is the geometric
mean of 29 individual workload ratios. Each SPEC runtime row is the geometric
mean of its selected test and train input ratios; the input column gives the
number of cases. Each SPEC compilation ratio covers one compiled program.
The separate-campaign and small-tests input qualifications stated for
Table~\ref{tab:optimization-progress} also apply here.

\begin{table}[htbp]
  \centering
  \caption{O2 performance of GPT-6-astra's selected candidates relative to original
  CompCert~3.17 ($=1$; lower is better). \chen{Rocq} denotes optimized CompCert~3.17
  (\texttt{2b22ad49}); Lean denotes optimized CompCert4Lean
  (\texttt{742233e}).}
  \label{tab:optimized-case-results}
  \begingroup
  \small
  \setlength{\tabcolsep}{3pt}
  \begin{tabular*}{\linewidth}{@{\extracolsep{\fill}}lrrrrr@{}}
    \toprule
    Benchmark & Inputs & \multicolumn{2}{c}{O2 program runtime}
      & \multicolumn{2}{c}{O2 compilation time} \\
    \cmidrule(lr){3-4}\cmidrule(l){5-6}
      & & \chen{Rocq} & Lean & \chen{Rocq} & Lean \\
    \midrule
    CompCert small-tests & 29 & 0.977 & 0.883 & 1.043 & 2.171 \\
    \midrule
    LBM & 2 & 0.998 & 0.967 & 1.025 & 1.371 \\
    Neutron & 4 & 1.002 & 0.964 & 1.058 & 1.880 \\
    TeaLeaf & 1 & 0.978 & 0.893 & 1.008 & 1.686 \\
    NAB & 1 & 0.982 & 0.898 & 1.063 & 1.812 \\
    Zstd & 1 & 0.897 & 0.831 & 1.058 & 1.502 \\
    SQLite & 1 & 0.925 & 0.905 & 1.552 & 2.466 \\
    \bottomrule
  \end{tabular*}
  \endgroup
\end{table}

The optimization traces suggest higher effective search throughput as one
explanation for the larger Lean gains. Within the evaluation windows---the
first 24 active hours in both environments---GPT-6-astra
produces \textbf{106 candidate variants in Lean versus 40 in \chen{Rocq}} that
complete full compiler and proof builds, a $2.65\times$ difference. We count
each development worktree with at least one successful full build once;
these counts include related parameter variants and combinations, rather
than only candidates selected for their performance. Recorded model usage
is broadly comparable: 453.79 million total tokens for Lean versus
498.12 million for \chen{Rocq}, including cached input, and 2.51 million versus
2.30 million output tokens, including reasoning. The higher realized
throughput gives the agent more opportunities to evaluate, combine, and
refine optimization candidates within the available time.

This result suggests greater practical capability of GPT-6-astra for
developing compilable optimizations in the Lean environment and offers a
plausible mechanism for its larger performance improvements. Completing a
full build is an intermediate milestone: both campaigns also contain
candidates that pass verification but do not improve measured performance.
Moreover, candidate variants can reuse existing proofs, and their number
does not measure independent optimization ideas or proof difficulty.
With one campaign per environment, the comparison does not isolate
language-specific model capability from differences in proof-library
support, repository architecture, build infrastructure, or search choices.
The evidence therefore supports an advantage in the observed optimization
process, while leaving the contributions of these factors unresolved.

DeepSeek~4.1~Flash did not fully follow the campaign prompts. Both runs
continued after encountering prior optimization metadata, despite an explicit
instruction to stop for review. Their reported performance also used
different evaluation procedures: the Lean aggregate omitted one of the 29
small workloads and used a different SPEC profile, while the Rocq report
emphasized SPEC test inputs measured with custom sampling. We therefore
remeasure every retained revision on the final-candidate lineages using the
GPT-6-astra evaluation protocol. Tables~\ref{tab:deepseek-optimization-progress}
and~\ref{tab:deepseek-optimized-case-results} use the same 29 small workloads
and ten SPEC inputs, sampling procedure, normalization, and workload
weighting as Tables~\ref{tab:optimization-progress}
and~\ref{tab:optimized-case-results}. The separate-campaign and small-tests
input qualifications stated there still apply. This reevaluation makes the
measurements more comparable, but cannot establish that the original
optimization searches were independent.

The DeepSeek search inventory in the first 24 active hours contains
48 successfully built Lean optimization worktrees and 76 named Rocq
compiler snapshots (72 distinct binaries); Lean reaches 57 including its
later continuation. Baselines and one diagnostic-only Lean worktree are
excluded. These counts differ from Astra's full compiler/proof-build
metric: most Lean builds targeted only the compiler, with full builds
recorded for two Lean worktrees (three with the continuation), while Rocq
reused one development worktree. Compilation sums translation-unit medians,
excluding linking. The block-layout checkpoint \texttt{671e092} fails
SQLite compilation at the block-layout validator.

\begin{table}[htbp]
  \centering
  \caption{DeepSeek~4.1~Flash optimization checkpoints, remeasured with the
  GPT-6-astra protocol. O2 time ratios use original CompCert~3.17 as the
  baseline ($=1$; lower is better). Each entry is the geometric mean over
  39 workloads: 29 small-tests programs and ten SPEC inputs. Within each
  compiler group and column, bold marks the best value and underlining
  the second best; ties share a rank and reference rows are not ranked.
  Dashes indicate unavailable aggregates because the block-layout
  checkpoint fails SQLite compilation.}
  \label{tab:deepseek-optimization-progress}
  \begingroup
  \small
  \setlength{\tabcolsep}{3pt}
  \begin{tabular*}{\linewidth}{@{\extracolsep{\fill}}llrr@{}}
    \toprule
    Compiler version & Change & \shortstack[r]{O2 program\\runtime}
      & \shortstack[r]{O2 compilation\\time} \\
    \midrule
    Original CompCert~3.17 & Baseline & 1.000 & 1.000 \\
    CompCert~3.17 \texttt{7074c50b} & Small-static inlining & 0.998 & \textbf{1.021} \\
    CompCert~3.17 \texttt{5f9087d3} & Wider small-static inlining & 0.991 & 1.058 \\
    CompCert~3.17 \texttt{aeb08ee0} & Called-once size bound & \underline{0.990} & \underline{1.042} \\
    CompCert~3.17 \texttt{c21ae16a} & Loop-weighted spill costs & \underline{0.990} & 1.059 \\
    CompCert~3.17 \texttt{6601adf5} & Spill-cost divisor refinement & \textbf{0.988} & 1.055 \\
    \midrule
    CompCert4Lean \texttt{3c06641} & Campaign baseline & 0.968 & 1.790 \\
    CompCert4Lean \texttt{671e092} & Block layout & \texttt{--} & \texttt{--} \\
    CompCert4Lean \texttt{b558b5e} & Block-layout fuel fix & 0.968 & 1.821 \\
    CompCert4Lean \texttt{44fd799} & Budgeted static-leaf inlining & \underline{0.958} & \underline{1.784} \\
    CompCert4Lean \texttt{97940a8} & Budgeted non-leaf inlining & 0.964 & 1.808 \\
    CompCert4Lean \texttt{5137462} & Called-once size bound & 0.962 & \textbf{1.747} \\
    CompCert4Lean \texttt{50d2c89} & Larger inlining budget & 0.963 & 1.785 \\
    CompCert4Lean \texttt{004762b} & Public-function inlining & \textbf{0.940} & 1.866 \\
    \bottomrule
  \end{tabular*}
  \endgroup
\end{table}

\begin{table}[htbp]
  \centering
  \caption{O2 performance of the best passing DeepSeek~4.1~Flash revisions
  under the GPT-6-astra evaluation protocol, relative to original
  CompCert~3.17 ($=1$; lower is better). Rocq denotes CompCert~3.17
  \texttt{6601adf5}; Lean denotes CompCert4Lean \texttt{004762b}. Runtime
  rows are geometric means of per-input median time ratios; each SPEC
  compilation ratio covers one compiled program. The normalization and
  small-tests input qualifications of Table~\ref{tab:optimized-case-results}
  also apply here.}
  \label{tab:deepseek-optimized-case-results}
  \begingroup
  \small
  \setlength{\tabcolsep}{3pt}
  \begin{tabular*}{\linewidth}{@{\extracolsep{\fill}}lrrrrr@{}}
    \toprule
    Benchmark & Inputs
      & \multicolumn{2}{c}{O2 program runtime}
      & \multicolumn{2}{c}{O2 compilation time} \\
    \cmidrule(lr){3-4}\cmidrule(l){5-6}
      & & Rocq & Lean & Rocq & Lean \\
    \midrule
    CompCert small-tests & 29 & 0.992 & 0.950 & 1.044 & 1.964 \\
    \midrule
    LBM & 2 & 1.000 & 0.974 & 1.091 & 1.293 \\
    Neutron & 4 & 1.003 & 0.902 & 1.012 & 1.708 \\
    TeaLeaf & 1 & 0.987 & 0.925 & 1.005 & 1.583 \\
    NAB & 1 & 0.993 & 0.965 & 1.028 & 1.523 \\
    Zstd & 1 & 0.863 & 0.811 & 0.998 & 1.369 \\
    SQLite & 1 & 0.928 & 0.867 & 1.781 & 2.495 \\
    \bottomrule
  \end{tabular*}
  \endgroup
\end{table}

Table~\ref{tab:kimi-optimization-progress} reports repeated measurements
of Kimi~K3.1's retained checkpoints. Every row covers all 29 small programs
and ten SPEC inputs, using the median of seven wall-time samples after one
warmup. Compilation sums three-sample median C-to-object times across
translation units and repeats each SPEC program's ratio for its input count.
All executions pass their output checks. Compiler order rotates each round
on one CPU. Intermediate checkpoints share a freshly measured
original-CompCert reference; the selected final revisions use the earlier
selected-candidate measurement batch. The Neutron repetitions were paused
and resumed with unchanged binaries and inputs, retaining completed samples.
Timing noise and the separate measurement batches limit comparisons of
small differences between checkpoints.

\begin{table}[htbp]
  \centering
  \caption{Repeated O2 measurements of Kimi~K3.1's retained checkpoints,
  relative to original CompCert~3.17 ($=1$; lower is better). Each entry is
  the geometric mean of 39 per-workload ratios using seven runtime samples
  and three compilation samples. The selected final rows use the same
  measurements as Table~\ref{tab:kimi-optimized-case-results}.}
  \label{tab:kimi-optimization-progress}
  \begingroup
  \small
  \setlength{\tabcolsep}{3pt}
  \begin{tabular*}{\linewidth}{@{\extracolsep{\fill}}llrr@{}}
    \toprule
    Compiler version & Change & \shortstack[r]{O2 program\\runtime}
      & \shortstack[r]{O2 compilation\\time} \\
    \midrule
    Original CompCert~3.17 & Baseline & 1.000 & 1.000 \\
    CompCert~3.17 \texttt{180e1bdd} & 24-node multi-call inlining & 0.991 & 1.041 \\
    CompCert~3.17 \texttt{f5ba9986} & 32-node multi-call inlining & 0.991 & 1.058 \\
    CompCert~3.17 \texttt{b530f8da} & Small leaves (selected) & 0.990 & 1.051 \\
    \midrule
    CompCert4Lean \texttt{3c06641} & Campaign baseline & 1.004 & 1.964 \\
    CompCert4Lean \texttt{29c6506} & 256-node static leaves & 0.994 & 2.001 \\
    CompCert4Lean \texttt{157b44b} & 8-node static non-leaves & 0.991 & 2.004 \\
    CompCert4Lean \texttt{4da16e9} & 176-node static leaf budget & 0.991 & 2.009 \\
    CompCert4Lean \texttt{2c791a5} & External single-call leaves & 0.992 & 2.005 \\
    CompCert4Lean \texttt{0d5a120} & Nine-case jump tables & 0.989 & 2.006 \\
    CompCert4Lean \texttt{65c4f34} & 8-node external non-leaves & 0.990 & 2.010 \\
    CompCert4Lean \texttt{33ce3a3} & 12-node external non-leaves & 0.991 & 2.012 \\
    CompCert4Lean \texttt{6810394} & 20-cost if-conversion & 0.990 & 2.001 \\
    CompCert4Lean \texttt{93dd165} & Balanced if-conversion (selected) & 0.979 & 2.001 \\
    \bottomrule
  \end{tabular*}
  \endgroup
\end{table}

The nine-case jump-table threshold is later reverted to seven cases;
\texttt{65c4f34} and subsequent checkpoints use seven. Restorations and empty
measurement commits with identical compiler sources are not separate rows.

Table~\ref{tab:kimi-optimized-case-results} reports the retained Kimi~K3.1
candidates: CompCert~3.17 \texttt{b530f8da} and CompCert4Lean
\texttt{93dd165}. Both extend inlining heuristics; the Lean candidate also
adjusts if-conversion costs. They reuse existing correctness proofs and
introduce no new verified pass. The completed Rocq crash-gap compensation
run retained the same selected implementation.

We remeasure both selected compilers and original CompCert~3.17 on all
29 small programs and ten SPEC inputs. All three compilers use identical
sources, inputs, and output checks, including the repeated 1\,MiB compression
input and \texttt{problem.dfg} for SPASS. Each workload has one warmup and
seven wall-time samples on one CPU; compiler order rotates each round.
Compilation uses three samples per translation unit. The shared reference
is measured in the same rounds, rather than taken from the archived release
measurements used for the earlier Lean campaigns. All timed executions pass
their output checks. These measurements replace the incomplete campaign
records and their different timing metrics.

\begin{table}[htbp]
  \centering
  \caption{Remeasured O2 performance of Kimi~K3.1's selected candidates,
  relative to original CompCert~3.17 ($=1$; lower is better). Rocq denotes
  \texttt{b530f8da}; Lean denotes \texttt{93dd165}. Runtime rows are geometric
  means of per-input median wall-time ratios. Both compilers use identical
  inputs and the same freshly measured reference. Each SPEC compilation
  ratio covers one compiled program.}
  \label{tab:kimi-optimized-case-results}
  \begingroup
  \small
  \setlength{\tabcolsep}{3pt}
  \begin{tabular*}{\linewidth}{@{\extracolsep{\fill}}lrrrrr@{}}
    \toprule
    Benchmark & Inputs
      & \multicolumn{2}{c}{O2 program runtime}
      & \multicolumn{2}{c}{O2 compilation time} \\
    \cmidrule(lr){3-4}\cmidrule(l){5-6}
      & & Rocq & Lean & Rocq & Lean \\
    \midrule
    CompCert small-tests & 29 & 0.994 & 0.977 & 1.031 & 2.068 \\
    \midrule
    LBM & 2 & 1.000 & 1.013 & 1.027 & 1.356 \\
    Neutron & 4 & 0.999 & 1.003 & 1.004 & 1.935 \\
    TeaLeaf & 1 & 1.002 & 1.001 & 1.003 & 2.030 \\
    NAB & 1 & 1.000 & 0.999 & 1.082 & 1.738 \\
    Zstd & 1 & 0.879 & 0.877 & 1.151 & 1.605 \\
    SQLite & 1 & 0.945 & 0.942 & 2.199 & 2.708 \\
    \bottomrule
  \end{tabular*}
  \endgroup
\end{table}

Compilation sums the three-sample median C-to-object times of each
translation unit, excluding linking and helper setup. Repeating each SPEC
program's compilation ratio for its input count gives the 39-workload ratios
1.051 and 2.001 in Table~\ref{tab:model-optimization-summary}.
For the selected candidates, omitting any one runtime round changes each
small-tests, SPEC, or combined aggregate by less than 0.1\% relative to the
reported value. Omitting one of the three compilation rounds changes the
compilation aggregate by up to 4.3\%. These checks describe sensitivity to
the retained samples; they do not establish confidence intervals.
The retained search records contain 33 named Rocq
experiments and 101 Lean experimental revisions reaching a passing
139-case O2 suite. Controls and final rechecks are excluded. These counts
describe a different milestone from Astra's full compiler/proof-build
counts; they count experiments with suite summaries retained in the
inspected inventory.

For Table~\ref{tab:model-optimization-summary}, Astra and DeepSeek SPEC
aggregates are reconstructed from rounded per-program ratios, weighted by
input count; Kimi uses unrounded input ratios. Compilation is the geometric
mean over the same 39 workloads, repeating each SPEC program's compilation
ratio for its inputs.
 
\clearpage
\section{Recorded Human Interactions}
\label{app:human-interventions}

This appendix supports Section~\ref{sec:human-effort} with the scope of the
recorded-message analysis and the full classification in
Table~\ref{tab:human-interactions}, followed by the three technical-direction
cases. We use the finalized version~1.6
classification. Technical direction denotes a supplied technical proposal,
not necessarily a correct diagnosis, a successful repair, or an override
of an agent recommendation.

\textbf{Corpus and classification.}
The analysis covers all 1,071 substantive human messages exchanged across
211 sessions during development from August~11 through RC1, the
\texttt{rc-0.0.1} release on September~24, 2026. Excluding 22
acknowledgments and greetings from the 1,093 human messages leaves the
1,071 substantive messages classified here. Machine-generated or replayed
content is not counted as human messages. Pre-release compiler validation
and execution-time work are included; subsequent optimization experiments,
manuscript editing, and maintenance are outside the development scope.
Each message has equal weight, regardless of length. An authored request
accompanying a copied agent plan counts once without attributing the
copied technical ideas to the human.

Three agents classified disjoint batches using a shared codebook and
calibration set, followed by additional review and coordinator
adjudication. Subsequent owner clarifications and review established the
final workflow boundary and the precedence rules below. A file assignment, proof-obligation
assignment, or branch/commit instruction alone is workflow, even when the
assigned work is technical. Questions about semantics, proof validity,
or algorithms are technical review/discussion; a supplied substantive
technical alternative is technical direction. The precedence rules place
each message in exactly one cell and do not rank expertise or importance.
The finalized corpus contains no unresolved primary categories.

Each message is classified by function as well as role. Information and
discussion seeks or exchanges information; advancing a plan directs work
within an existing plan; establishing or changing a plan sets or revises
the plan. Both dimensions are hierarchical and mutually exclusive. For
function, establishing or changing a plan takes precedence over advancing
it, which takes precedence over information and discussion. For role,
human technical direction takes precedence over technical review and
discussion, which takes precedence over workflow. A concrete implementation
alternative therefore counts as technical direction even when motivated
by keeping work moving. Table~\ref{tab:human-interactions} reports all nine
combinations and their totals; Figure~\ref{fig:human-interaction-roles}
aggregates the columns.

\begin{table}[htbp]
  \centering
  \caption{Recorded human interactions during migration. Each of 1,071
  substantive messages from 211 sessions occupies one cell. Counts are
  messages, not tokens; row and column totals include all substantive messages.}
  \label{tab:human-interactions}
  \begingroup
  \small
  \setlength{\tabcolsep}{3pt}
  \begin{tabular*}{\linewidth}{@{\extracolsep{\fill}}lrrrr@{}}
    \toprule
    Function & Workflow & \shortstack{Technical review\\/ discussion}
      & \shortstack{Human technical\\direction} & Total \\
    \midrule
    Information \& discussion & 387 & 70 & 0 & 457 \\
    Advance plan & 326 & 39 & 0 & 365 \\
    Establish/change plan & 238 & 8 & 3 & 249 \\
    \midrule
    Total & 951 & 117 & 3 & 1,071 \\
    \bottomrule
  \end{tabular*}
  \endgroup
\end{table}

The 117 review/discussion messages are not all approvals. Across the 120
messages assigned to either technical role, the records contain 45
explicit approvals and 77 inquiry/discussion/investigation incidences,
with four messages in both subcategories; one technical-direction message
also contains an inquiry. The three direction messages comprise two
contributions and one confirmed override. These overlapping subcounts do
not partition the review column.

\textbf{Tentative proposition repair.}
During definition review, an agent identified propositions replaced by
\texttt{True} and explained that the exact source statements needed to be
restored. The human proposed dependency-ordered repair across twelve
modules and tentatively suggested replacing the trivial propositions with
\texttt{sorry}. The agent corrected this suggestion: workers should
reconstruct the exact propositions and leave their proofs absent, rather
than admit arbitrary proposition definitions. It then developed the
dependency plan, which the human authorized. The finalized classification
counts the concrete substitution suggestion as a technical contribution,
while dependency scheduling is workflow. This records a tentative
proposal corrected by the agent, not an independently supplied human
diagnosis or the successful repair strategy.

\textbf{Conditional allocator fallback.}
While planning a runnable prototype, the agent recommended reusing and
repairing the existing spill allocator. The human added a conditional
alternative: switch to CompCert's allocator if repairing the prototype
proved difficult, so that allocation would not block end-to-end testing
within the existing verification boundary. This permitted the original
prototype-first approach and is classified as a contribution, not an
override. The agent supplied the concrete bridge design, a one-attempt,
30-minute repair bound, and a fallback to upstream OCaml allocation whose
proposals would pass through the existing Lean validator. The retained
completion report states that three smoke tests passed after one local
repair and the spill allocator remained the default; it does not report
using the fallback.

\textbf{Change toward a CompCert-style allocator.}
In a later investigation in the same allocator-development episode, an
agent recommended local repairs to builtin-specific placement in the
prototype and additional pre-expansion checks. In an explicitly linked
follow-up session, the human instead proposed implementing an allocator
matching CompCert's approach and asked about its correctness boundary.
This is the one confirmed override in the finalized classification: the
record connects the earlier recommendation to the subsequent contrary
direction. The agent explained the upstream allocation stages and the
existing RTL-to-LTL validator, and the human authorized implementation.
The allocation algorithm, bridge details, and validation explanation were
supplied by the agent, not by the human's directional proposal. A later
agent report records a Lean-native iterated-register-coalescing allocator
with iterative spilling and x86 placement checks, with all 139 applicable
tests passing. These are retained completion reports, not independent
rechecks performed by the message analysis; the record establishes a
change of approach, not its necessity or superiority.

\textbf{Interpretation.}
Classification review provides agent-based consistency checking with owner
adjudication. Message shares measure neither human working hours nor
reading, offline reasoning, expertise, or causal importance. Even a rare
direction may be consequential. The separate prompt-token total in
Section~\ref{sec:human-effort} is not the denominator of this analysis.
 
\end{document}